\documentclass[aps,prd,preprint,amsmath,citeautoscript,longbibliography,nofootinbib]{revtex4-1}
\usepackage[utf8]{inputenc}
\usepackage{bm}
\usepackage{amsmath,mathrsfs,amsfonts}
\usepackage{graphicx,afterpage}
\usepackage{subcaption}
\usepackage{amsmath,latexsym}
\usepackage{color,soul}
\usepackage{float}
\usepackage[section]{placeins}
\usepackage{silence}
\begin{document}
\title{New class of exact solutions to Einstein-Maxwell-dilaton theory on {four}-dimensional Bianchi type IX geometry}
\author{Bardia H. Fahim}
\email{bardia.fahim@usask.ca}
\author{Masoud Ghezelbash}
\email{Corresponding author, masoud.ghezelbash@usask.ca}
\affiliation{Department of Physics and Engineering Physics, University of Saskatchewan, Saskatoon SK S7N 5E2, Canada}
\date{\today}
\begin{abstract}
We construct new classes of cosmological solution to the five dimensional Einstein-Maxwell-dilaton theory, that are non-stationary and almost conformally regular everywhere. The base geometry for the solutions is the four-dimensional Bianchi type IX geometry.
In the theory, the dilaton field is coupled to the electromagnetic field and the cosmological constant term, with two different coupling constants. We consider all possible solutions with different values of the coupling constants, where the cosmological constant takes any positive, negative or zero values.
In the ansatzes for the metric, dilaton and electromagnetic fields, we consider dependence on time and two spatial directions. We also consider a special case of the Bianchi type IX geometry, in which the geometry reduces to that of Eguchi-Hanson type II geometry and \textcolor{black} {find a more general solution} to the theory.
\end{abstract}
\maketitle

\section{Introduction}
One of the main aims of gravitational physics is to find the exact solutions to the Einstein gravity in the presence of matter fields in different dimensions. The ideas of higher dimensional gravity and dimensional compactification \textcolor{black}{have} been explored \textcolor{black}{extensively} in different articles \cite{freund1980dynamics,kaluza2018unification}.
 Moreover, a better insight about the holography requires to construct and understand the exact solutions to the Einstein gravity
in asymptotically de-Sitter and Anti-de-Sitter spacetimes \cite{guica2009kerr}. The solutions to the Einstein theory in the background of different matter fields such as Maxwell field, dilaton field and NUT charges \textcolor{black}{are} explored in \cite{gouteraux2012holography, mahapatra2018time, ghezelbash2015cosmological}. The relevant solutions can be found in the compactification of M-theory in generalized Freund-Rubin theory \cite{torii2003cosmological}. The applications and properties of the Einstein-Maxwell-dilaton theory can be found in different areas such as slowly rotation black holes \cite{stetsko2019slowly,sheykhi2010higher}, {\color{black} topological charged hairy black holes \cite{add1},} cosmic censorship \cite{goulart2018violation}, gravitational radiation \cite{julie2018gravitational} and hyperscaling violation \cite{li2017hyperscaling}. {\color{black} Moreover, the variation of the standard Einstein-Maxwell-dilaton theory with two extra vector fields, where one field supports the non-trivial topology, and the second field supports states with the finite charge density, was considered in \cite{add2}.  In the context of the generalized  Einstein-Maxwell-dilaton theory, the authors found a new class of charged black holes with hyper-scaling violating asymptotics and non-trivial horizon topology, for arbitrary Lifshitz exponent and a hyper-scaling violation parameter \cite{add2}.}

In this article, we explore the exact solutions to the five-dimensional Einstein-Maxwell-dilaton theory with two coupling constants and a cosmological constant.
We find the exact solutions to the five-dimensional Einstein-Maxwell-dilaton theory where the dilaton field is coupled to both the electromagnetic field and the cosmological constant with two different coupling constants. We find expressions for the Maxwell field and dilaton field. We also find the cosmological constant in terms of the coupling constant and show that it can only take specific numbers. We find a relation between the coupling constants and an extra constraint which limits the coupling constant to certain numbers.  We show that the solutions cannot be uplifted to a higher dimensional Einstein gravity or Einstein-Maxwell theory with the cosmological constant. We also find the exact solutions to the theory for the case where the coupling constants are equal and non-zero, and where the coupling constants are both equal to zero. Considering the latter case, the Einstein-Maxwell-dilaton theory reduces to the Einstein-Maxwell theory in the presence of the cosmological constant. We discuss the properties of the non-stationary spacetimes and show that our exact non-trivial solutions satisfy all the field equations.

{\textcolor{black} {We should emphasize our exact solutions exist only in five dimensional Einstein-Maxwell-dilaton theory with two coupling constants and a cosmological constant. We consider ansatzes for the metric, the Maxwell field and the dilaton which perfectly lead to exact solutions in three different cases: I) the coupling constants are different, II) the coupling constants are non-zero and equal and III) the coupling constants are zero.  We explicitly show in appendix A, that our ansatzes work only in five dimensions. To find the higher dimensional solutions to the Einstein-Maxwell-dilaton theory, we might consider adding other fields, such as extra vector fields to the standard Einstein-Maxwell-dilaton theory \cite{add2}, to support the existence of the solutions, in dimensions greater than five. }}

The article is organized as follows:\newline
In section 2, we find the exact solutions to the Einstein-Maxwell-dilaton theory based on the Bianchi type IX metric. The dilaton field is coupled to the electromagnetic field and the cosmological constant with two different coupling constants. We consider specific ansatzes for the five-dimensional spacetime, Maxwell field and the dilaton field. We solve all the equations of motion and find the metric functions for the five-dimensional spacetime. We find the c-function for the spacetime and discuss the properties of the spacetime. Moreover, we find a relation between the coupling constants. 

In section 3, we consider the coupling constants to be equal to each other and find the exact solutions to the Einstein-Maxwell-dilaton theory. We consider two different cases, where the coupling constant are equal to each other and are non-zero, and where the coupling constants are zero. Each case needs a different ansatz for the metric functions. 

In section 4 we present a combination of the solutions based on the four-dimensional Eguchi-Hanson space, which is a subspace of the Bianchi type IX geometry. We verify the ansatzes and show that our assumptions satisfy all the equations of Einstein, Maxwell and dilaton. 

In section 5, we discuss the uplifting of the exact solutions to higher dimensional theories such as Einstein gravity in a higher dimension and Einstein-Maxwell theory with a cosmological constant. We show that our solutions cannot be found from the compactification of these theories. Moreover, we calculate the Kretschmann invariant and discuss the singularities of the spacetime. 

{\textcolor{black} {We end the article with concluding remarks and three appendices. In appendix A, we show that the exact solutions exist only in five dimensions. In appendix B, we present the classification of the Bianchi spaces and show Bianchi type IX possess the maximal symmetry between all types of Bianchi spaces. Hence Bianchi type IX is the best space to be uplifted to the higher dimensional Einstein-Maxwell-dilaton theory. In appendix C, we present explicitly the Maxwell field equations and their solutions.}}

\section{Exact Solutions to the Einstein-Maxwell-dilaton theory, Based on Bianchi type IX geometry, with two different coupling constants $a$ and $b$}

We consider the cosmological Einstein-Maxwell-dilaton (EMD) theory, where the dilaton field interacts with both the cosmological constant and the electromagnetic field, with two different coupling constants. 
The action for the theory in presence of the cosmological constant $\Lambda$ is given in $N+1$ dimensions by \cite{maki1993multi},
\begin{equation}
    S=\int d^{N+1}x\sqrt{-g}\{R-\frac{4}{N-1}(\nabla\phi)^2-e^{-4/(N-1)a\phi}F^2-e^{4/(N-1)b\phi}\Lambda\}, \label{action}
\end{equation} 
 where $R$ represents the curvature scalar, $g=det[g_{\mu\nu}]$ and $F_{\mu \nu}$ is the electromagnetic field strength \cite{hobson2006general,misner2017gravitation}. Moreover, in equation (\ref{action}), $\phi$ is the dilaton field and $a$ and $b$ are two arbitrary coupling constants.

We find the Maxwell field equations in $N+1$ dimensions, by varying the given action in equation (\ref{action}) with respect to the electromagnetic gauge field $A_\mu$ \cite{maki1993multi},
\begin{equation}
    M_\mu\equiv \nabla^\nu(e^{-4/(N-1)a\phi}F_{\mu \nu})=0. \label{maxwell}
\end{equation}

By varying the action (\ref{action}) with respect to the dilaton field, we find the dilaton field equations in $N+1$ dimensions as,
\begin{equation}
    D\equiv \nabla^2\phi-\frac{b}{2}e^{4/(N-1)b\phi}\Lambda +\frac{a}{2}e^{-4/(N-1)a\phi}F^2=0. \label{dilaton}
\end{equation}

Moreover, varying the Einstein-Maxwell-dilaton action (\ref{action}) with respect to the metric tensor $g_{\mu \nu}$ leads to the Einstein field equations in $N+1$ dimensions \cite{maki1993multi},
\begin{eqnarray}
    \varepsilon_{\mu \nu} &\equiv& R_{\mu \nu}-\frac{1}{2}g_{\mu \nu} R-\frac{4}{N-1}\{\nabla_\mu\phi\nabla_\nu\phi-\frac{1}{2}g_{\mu \nu}(\nabla\phi)^2\}\nonumber\\
    &-& \, e^{\frac{-4a\phi}{N-1}}\{2F_{\mu \lambda} F_\nu^\lambda-\frac{1}{2}g_{\mu \nu} F^2\}-\frac{1}{2}e^{\frac{4b\phi}{N-1}}g_{\mu \nu}\Lambda{=0}.\label{ein}
\end{eqnarray}
{\color{black} In this article, we consider the four-dimensional Bianchi type IX geometry, as the background metric to be uplifted to the Einstein-Maxwell-dilaton theory. The Bianchi type IX geometry is included in the classification of the homogeneous spaces, which was done in 1897 by Bianchi, and later on used in cosmology by Lifschitz, Belinski and Khalatnikov \cite{belinskii1969nature}.  In appendix B, we present the classification of the homogeneous spaces, which leads to choosing the Bianchi type IX geometry (between all Bianchi type I,$\cdots$, IX geometries) possessing the maximal symmetry.

Among all the different four-dimensional Bianchi type geometries, we consider the Bianchi type IX geometry, which not only posses the maximal symmetry (between all Bianchi type I,$\cdots$, IX geometries), but also is a self-dual and asymptotically locally Euclidean space. The self-duality and maximal symmetry of the Bianchi type IX, enable us to find the exact solutions by uplifting the Bianchi type IX geometry into the Einstein-Maxwell-dilaton theory. 

Moreover, the other reason for choosing Bianchi type IX is that only this space (between all Bianchi type I,$\cdots$, IX spaces) reduces exactly to the well-known spaces, such as Taub-NUT, Eguchi-Hanson type I and type II and Atiyah-Hitchin geometries, in some  appropriate limits.  

The latter spaces have been studied extensively in different theories of the gravitational physics, including uplifting them to the higher-dimensional extensions of gravity, such as string cosmological model in string theory \cite{bali2001bianchi}, loop quantum cosmology \cite{wilson2010loop}, supergravity \cite{ghezelbash2008supergravity}, and M-branes \cite{ghezelbash2006bianchi}. In fact, the exact solutions to the Einstein-Maxwell-dilaton theory have been constructed for the embedded Eguchi-Hanson type II geometry \cite{ghezelbash2017new}. In this article, we even find more general exact solutions for uplifting the Eguchi-Hanson type II into the Einstein-Maxwel-dilaton theory, by using the exact solutions for the embedded Bianchi type IX geometry and then reducing them to the Eguchi-Hanson type II geometry.}


The triaxial Bianchi type IX metric is given by \cite{ghezelbash2008supergravity},
\begin{eqnarray}
     ds^2_{tr. BIX} &=& \frac{dr^2}{\sqrt{J(r)}}+\frac{r^2}{4}\sqrt{J(r)}\{\frac{(d\psi+\cos{\theta}d\phi)^2}{1-\frac{a_1^4}{r^4}} \nonumber\\
     &+& \, \frac{(-\sin{\psi}d\theta+\cos{\psi}\sin{\theta}d\phi)^2}{1-\frac{a_2^4}{r^4}}+\frac{(\cos{\psi}d\theta+\sin{\psi}\sin{\theta}d\phi)^2}{1-\frac{a_3^4}{r^4}}\}. \label{trbix2}
\end{eqnarray}
The metric function $J(r)$ in the triaxial Bianchi type IX metric (\ref{trbix2}) is,
\begin{equation}
      J(r)=(1-\frac{a_1^4}{r^4})(1-\frac{a_2^4}{r^4})(1-\frac{a_3^4}{r^4}), \label{J}
\end{equation}
where $a_1$, $a_2$ and $a_3$ are three parameters. We choose these parameters to be $a_1=0$, $a_2=2kc$ and $a_3=2c$, where $c>0$ is a constant and $k$ belongs to the interval $0\leq k \leq 1$
\cite{ghezelbash2008supergravity}. The periodicity for the angles $\theta$, $\psi$ and $\phi$ are $\pi$, $4\pi$ and 2$\pi$, respectively. We note the coordinate $r$ should be $r\geq a_3$, otherwise the metric function $J(r)$ in equation (\ref{J}) becomes negative and therefore the metric (\ref{trbix2}) would contain imaginary parts.
We rewrite the Bianchi type IX metric in a more compact way in terms of the Maurer-Cartan one-forms $\sigma_i$,
 \begin{equation}
     ds^2_{B.IX}=\frac{dr^2}{{J(r)}^{1/2}}+\frac{r^2}{4}{J(r)}^{1/2}(\frac{\sigma^2_1}{1-\frac{a^2_1}{r^4}}+\frac{\sigma^2_2}{1-\frac{a^2_2}{r^4}}+\frac{\sigma^2_3}{1-\frac{a^2_3}{r^4}}).\label{BIX2}
 \end{equation}
The Maurer-Cartan one-forms are given by,
 \begin{equation}
     \sigma_1=d\psi+\cos{\theta}d\phi,
 \end{equation}
 \begin{equation}
     \sigma_2=-\sin{\psi}d\theta+\cos{\psi}\sin{\theta}d\phi,
 \end{equation}
 \begin{equation}
     \sigma_3=\cos{\psi}d\theta+\sin{\psi}\sin{\theta}d\phi.
 \end{equation}
 
We note that the Bianchi type IX geometry contains two well-known spaces, namely Eguchi-Hanson type I and type II in some appropriate limits. Considering $k=0$ in equation (\ref{trbix2}), the Bianchi type IX metric reduces to the Eguchi-Hanson type I metric, which is given by \cite{ghezelbash2008supergravity},
\begin{equation}
    ds_{EH. I}^2=\frac{dr^2}{f(r)}+\frac{r^2}{4}f(r)\{d\theta^2+\sin^2{\theta}d\phi^2\}+\frac{r^2}{4f(r)}(d\psi+\cos{\theta}d\phi)^2,\label{EH}
\end{equation}
where the metric function $f(r)$ is,
\begin{equation}
    f(r)=\sqrt{1-\frac{16c^4}{r^4}}.\label{EHf}
\end{equation}
Moreover, by choosing $k=1$ in equation (\ref{trbix2}), the Bianchi type IX metric reduces to the Eguchi-Hanson type II metric, which is given by \cite{singh20073}, 
\begin{equation}
    ds_{EH. II}^2=\frac{dr^2}{f(r)^2}+\frac{r^2f(r)^2}{4}(d\psi+\cos{\theta}d\phi)^2+\frac{r^2}{4}(d\theta^2+\sin^2{\theta}d\phi^2), \label{EHIIBB}
\end{equation}
where $f(r)$ is given in (\ref{EHf}).
The exact solutions to the Einstein-Maxwell-dilaton theory based on the latter metric, are reviewed later in section 4.

{\color{black}  To embed the four-dimensional Bianchi type IX metric into the Einstein-Maxwell-dilaton theory, $N$ should be greater than or  equal to 4 in equation (\ref{action}). Of course, not only $N=4$ gives the simplest theory to be explored, but also we can not find any exact solutions, where $N \geq 5$.  In fact, as we show in appendix A, for $N=5$, we can not consistently satisfy all the field equations, except by choosing trivial metric functions. So, though we presented the action (\ref{action}) for the Einstein-Maxwell-dilaton theory in $N+1$-dimensions, we only consider $N=4$ in what follows.
}The action for the Einstein-Maxwell-dilaton theory in presence of the cosmological constant, can be written in five-dimensions (4+1 dimensions) as,  
\begin{equation}
    S=\int d^5x \sqrt{-g}\{R-\frac{4}{3}(\nabla\phi)^2-e^{-4/3a\phi}F^2-e^{4/3b\phi}\Lambda\}, \label{action5}
\end{equation}
where the dilaton field is coupled to the Maxwell field (by the coupling constant $a$) and to the cosmological constant (by the coupling constant $b$). In the action (\ref{action5}), $R$ is the Ricci scalar, $\phi$ is the massless dilaton field, $F_{\mu \nu}$ is the electromagnetic tensor and $\Lambda$ represents the cosmological constant.

First, we consider the most general case, where the coupling constants are $a\neq b$ and non-zero. The application of this case can be found in the generalized Freund-Rubin compactification \cite{torii2003cosmological}.  

Varying the Einstein-Maxwell-dilaton action (\ref{action5}) with respect to the electromagnetic gauge field $A_\mu$ leads to the Maxwell field equations in five-dimensions,
\begin{equation}
     M_\mu=\nabla^\nu(e^{-4/3a\phi}F_{\mu \nu})=0.
\end{equation}
Moreover, we find the dilaton field equation in five-dimensions, by varying the action (\ref{action5}) with respect to the dilaton field,
\begin{equation}
     D=\nabla^2\phi+\frac{a}{2}e^{-4/3a\phi}F^2-\frac{b}{2} e^{4/3b\phi}\Lambda=0.
\end{equation}
The Einstein field equations can be found in five-dimensions, by varying the action (\ref{action5}) with respect to the metric tensor $g_{\mu \nu}$ \cite{ghezelbash2015cosmological},
\begin{equation}
     \varepsilon_{\mu \nu}=R_{\mu \nu}-\frac{2}{3}\Lambda g_{\mu \nu} e^{4/3b\phi}-(F_\mu^\lambda F_{\nu \lambda}-\frac{1}{6}g_{\mu\nu} F^2)e^{-4/3a\phi}-\frac{4}{3}\nabla_\mu\phi\nabla_\nu\phi=0.
\end{equation}

\begin{figure}[H]
\centering
\includegraphics[scale=0.50]{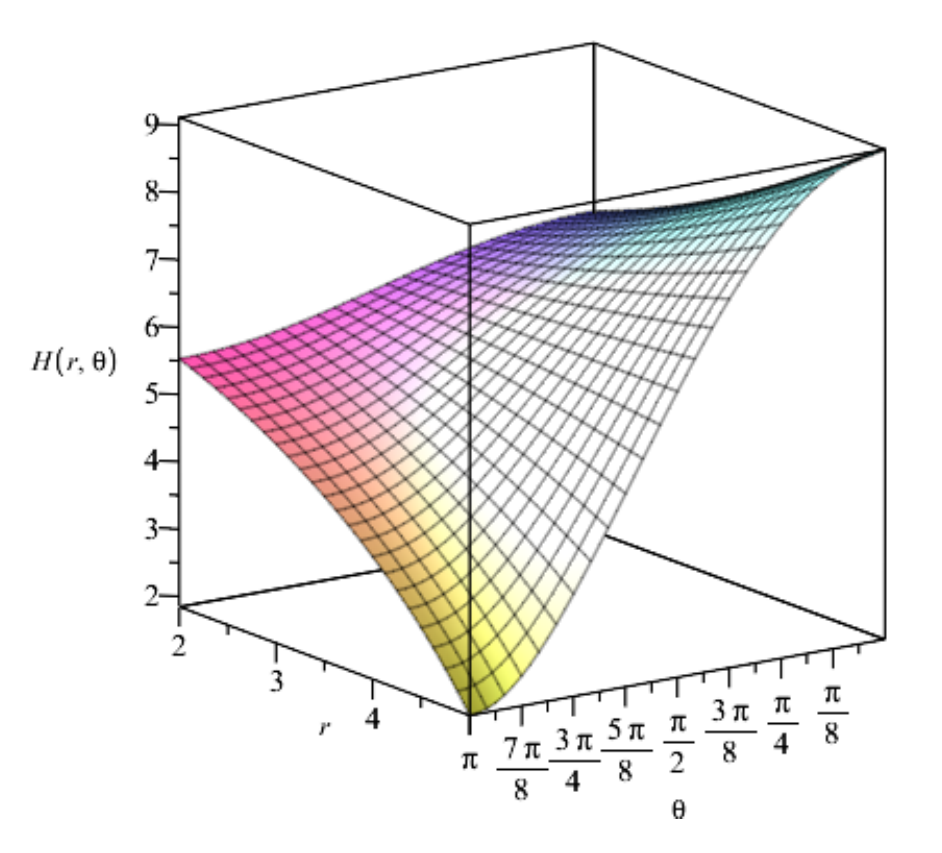}
\caption{The behaviour of the metric function $H(r,\theta)$ with respect to the coordinates $r$ and $\theta$, where we consider the constants as $j_+=0.5$, $j_-=15$, $c=1$ and $a=1$.} 
\label{fig:hab}
\end{figure}

We consider an ansatz for the five-dimensional metric as,
\begin{equation}
    ds_5^2=-\frac{1}{H^2(r,\theta)}dt^2+R^2(t)H(r,\theta)ds_{B. IX}^2. \label{st}
\end{equation}
In the ansatz (\ref{st}), $ds_{B.IX}^2$ represents the four-dimensional Bianchi type IX metric given by equation (\ref{trbix2}), and $H(r,\theta)$ and $R(t)$ are two metric functions. We consider the electromagnetic gauge field and the dilaton field in terms of the metric functions $H(r,\theta)$ and $R(t)$ as,
\begin{equation}
  A_t(t,r,\theta)=\alpha R^M(t)H^E(r,\theta)  , \label{A}
\end{equation}
\begin{equation}
    \phi(t,r,\theta)=-\frac{3}{4a}\ln{(H^L(r,\theta)R^W(t))}, \label{dil}
\end{equation}
where $\alpha$, $M$, $E$, $L$ and $W$ are arbitrary constants. According to the considered ansatz (\ref{A}), the only non-zero component of the electromagnetic gauge field is the $t$ component, which is a function of time and spatial coordinates $r$ and $\theta$. We find the electromagnetic field strength $F_{\mu \nu}$, as given by,
\begin{equation}
  F_{\mu \nu}=
  \left[ {\begin{array}{ccccc}
   0 & \alpha H^EE(\frac{\partial H}{\partial r})R^M/H & \alpha H^EE(\frac{\partial H}{\partial \theta})R^M/H & 0 & 0 \\
   - \alpha H^EE(\frac{\partial H}{\partial r})R^M/H & 0 & 0 & 0 &0 \\
   -\alpha H^EE(\frac{\partial H}{\partial \theta})R^M/H & 0 & 0 & 0 & 0 \\
   0 & 0 & 0 & 0 & 0 \\
   0 & 0 & 0 & 0 & 0 \\
  \end{array} } \right].
\label{eq:F}
\end{equation}
Based on the considered ansatzes (\ref{st}), (\ref{A}) and (\ref{dil}), the $r$ component of the Maxwell field equations $M^r$ becomes,
\begin{eqnarray}
    M^r&=&\frac{-1}{4a}\sqrt{\frac{(2ck-r)(2ck+r)(4c^2k^2+r^2)(2c-r)(2c+r)(4c^2+r^2)}{r^8}}\nonumber\\
   &\times& \, H^E(r,\theta)(\frac{\partial H}{\partial r})R^M(t)E\alpha(\frac{\partial R}{\partial t})(4aW+4Ma+8a),\label{Mr}
\end{eqnarray}
where $c$ and $k$ are the constants that appear in the Bianchi type IX metric (\ref{trbix2}).  From (\ref{Mr}), we find that the constants $M$ and $W$ satisfy,
 \begin{equation}
 M+W=-2. \label{MW}
 \end{equation}
In Appendix C, we show that the Maxwell's components $M^\phi$, $M^\psi$ and $M^\theta$, lead to the same constraint on the constants $M$ and $W$, as in equation (\ref{MW}).  
Moreover, from the Einstein’s field equation $\varepsilon_{t r}$, 
\begin{equation}
   \varepsilon_{t r}=\frac{-3}{4}\frac{(\frac{\partial R}{\partial t})(\frac{\partial H}{\partial r})(LW+4a^2)}{H(r,\theta)a^2R(t)},
\end{equation}
we find the following relation between the constants $L$ and $W$,
\begin{equation}
    LW+4a^2=0.
\end{equation}
From the $\varepsilon_{r \theta}$ component of Einstein equations,
\begin{equation}
   \varepsilon_{r \theta}= 4\alpha^2(H^E(r,\theta))^2E^2(R^M(t))^2H^2(r,\theta)a^2R^W(t)H^L(r,\theta)-3L^2-6a^2=0,\label{rth}
\end{equation}
and comparing it with other equations, we find the constants $M$, $E$ and $\alpha$ in (\ref{A}) as $M=2$, $E=-1-\frac{a^2}{2}$, and $\alpha^2=\frac{3}{a^2+2}$. The constants $L$ and $W$ in (\ref{dil}) are found to be $L=a^2$ and $W=-4$. Therefore, equation (\ref{rth}) satisfies as $\varepsilon_{r\theta}=0$. We present the $t$ component of the Maxwell’s equation $M^t$ in appendix C. Analyzing the equation, we find the solutions for the metric function $H(r,\theta)$ as follow,
\begin{equation}
H(r,\theta)=(j_+r^2\cos{\theta}+j_-)^{\frac{2}{a^2+2}}, \label{H}
\end{equation}
where $j \pm$ are two arbitrary constants. We present the behaviour of the metric function $H(r,\theta)$ with respect to the coordinates $r$ and $\theta$ in figure \ref{fig:hab}, where we set $j_+=0.5$, $j_-=15$ and $a=1$.
Through the other Einstein and Maxwell field equations, we find the solutions for the metric function $R(t)$ and the cosmological constant $\Lambda$, as,
\begin{equation}
    R(t)=(\eta t+\vartheta)^{a^2/4},\label{RR}
\end{equation}
\begin{equation}
    \Lambda=\frac{3}{8}\eta^2a^2(a^2-1), \label{cosmoc}
\end{equation}
where $\eta$ and $\vartheta$ are arbitrary constants and $a$ is the coupling constant. We find a relation between the coupling constants as,
\begin{equation}
    ab=-2.\label{abm2}
\end{equation}

Moreover, according to the five-dimensional metric in equation (\ref{st}), we notice that the metric function $H(r,\theta)$ should be a real and positive function. Therefore, we find the following constraint on the coupling constant $a$,
\begin{equation}
    a^2+2=2n+1, \label{constraint}
\end{equation}
where   $n\in \mathbb{N}$.  We notice that by choosing the coupling constant $a=1$ and $a=\sqrt{3}$, the Einstein-Maxwell-dilaton action (\ref{action}) leads to the low-energy effective action for heterotic string theory and Kaluza-Klein reduction of five-dimensional Einstein gravity, respectively \cite{rocha2018self}.

Furnished with all the results in equation (\ref{st}), by calculating the Ricci scalar and the Kretschmann invariant, we find that they diverge on the hyper-surfaces  $H(r,\theta)=0$ and $R(t)=0$. We should note that the same type of singularities exists for the supergravity solutions (in more than four-dimensions) \cite{ghezelbash2006bianchi}, which can be avoided with considering more spatial coordinates in the metric functions.

Moreover, by restricting the constants $\eta$ and $\vartheta$ in the metric function (\ref{RR}), the singularity at $R(t)=0$ can be removed,
\begin{equation}
    \eta \geq 0,
\end{equation}
\begin{equation}
    \vartheta>0.
\end{equation}
Considering the found relation between the coupling constants (\ref{abm2}), we rewrite the action (\ref{action5}) as,
\begin{equation}
      S=\int d^5x \sqrt{-g}\{R-\frac{4}{3}(\nabla\phi)^2-e^{\frac{-4a}{3}\phi}F^2-e^{\frac{-8}{3a}\phi}\Lambda\}.
      \label{eq:actionab}
\end{equation}

\begin{figure}[ht]
    \centering
    \includegraphics[scale=0.50]{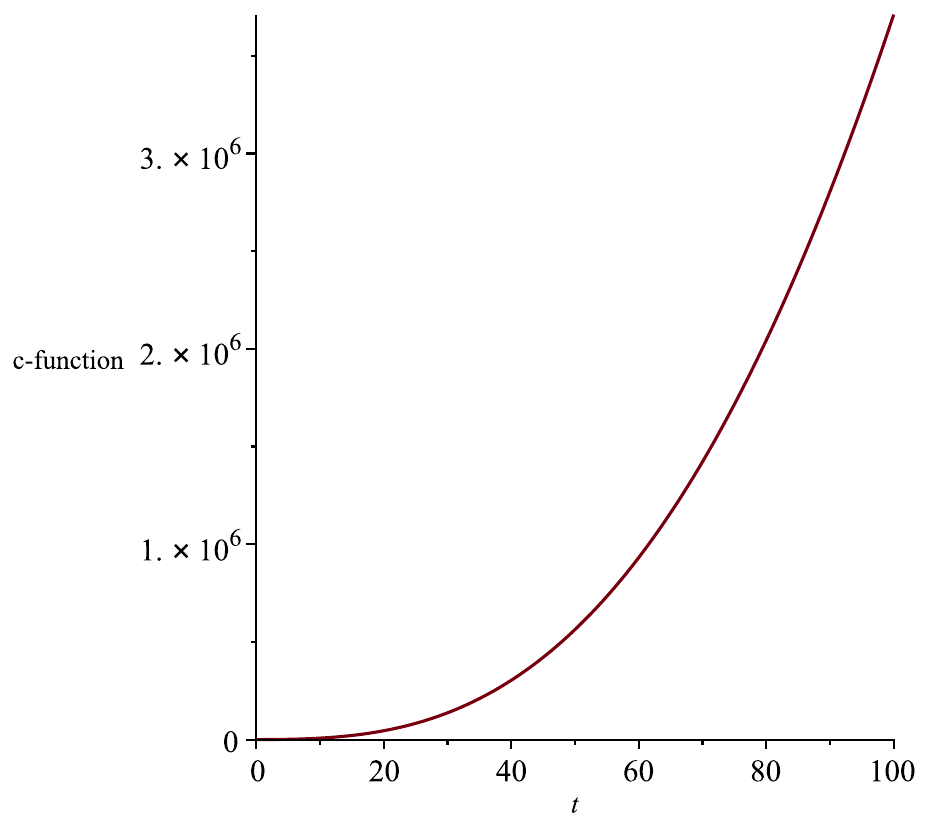}
    \captionsetup{justification=raggedright, singlelinecheck= false}
    \caption{The behaviour of the c-function for the five-dimensional spacetime, where the coupling constants are not equal.}
    \label{fig:cab1}
\end{figure}

\begin{figure}[H]
    \centering
    \includegraphics[scale=0.50]{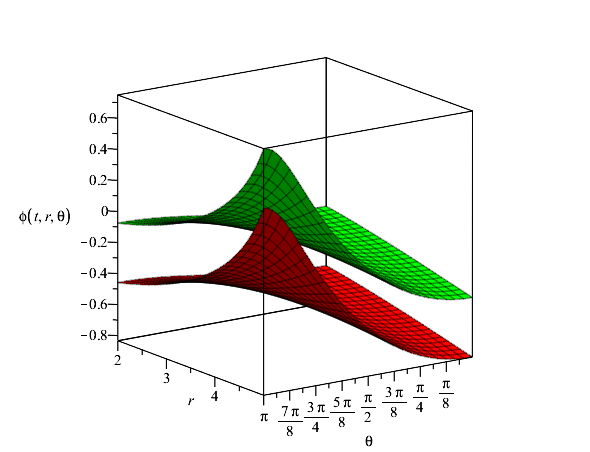}
    \caption{The behaviour of the dilaton field $\phi(t,r,\theta)$ for two different time slices, $t=1$ and $t=3$ (which are the lower and the upper surface, respectively), as a function of the coordinates $r$ and $\theta$. We  consider specific values for the set the constants $j_+=0.5$, $j_-=15$, $a=1$, $\eta=1$ and $\vartheta=2$.}
    \label{fig:phab}
\end{figure}

\begin{figure} [H]
    \centering
    \includegraphics[scale=0.50]{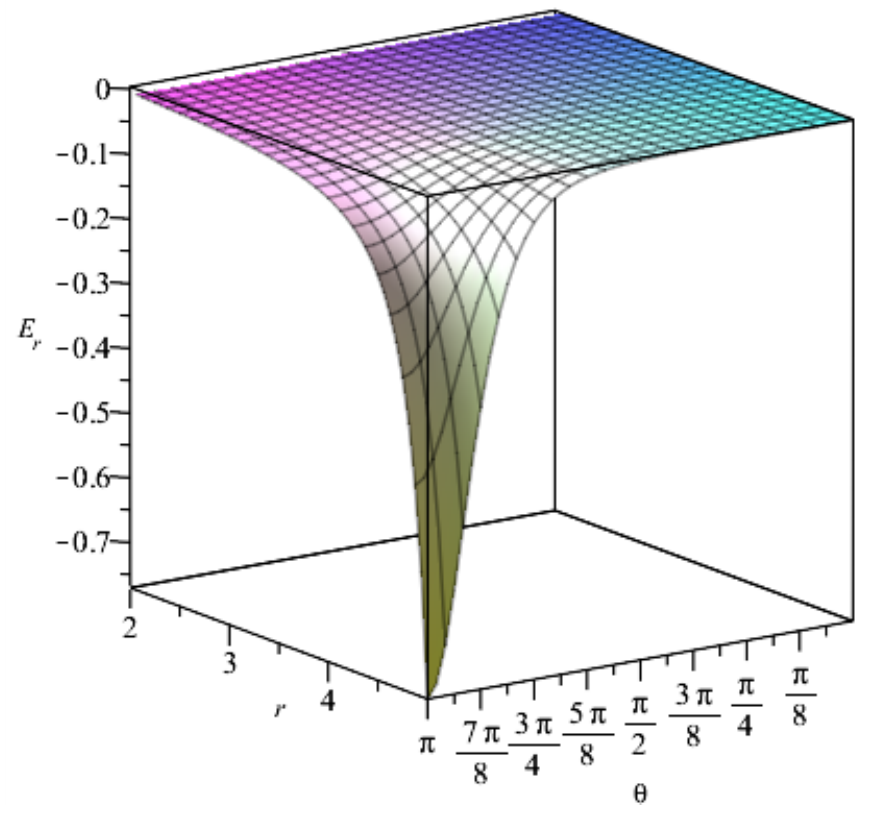}
    \caption{The behaviour of the $r$-component of the electric field for $t=1$, as a function of the coordinates $r$ and $\theta$, where we set $j_+=0.5$, $j_-=15$, $a=1$, $\eta=1$, $\vartheta=2$, $c=1$. }
    \label{fig:erab}
\end{figure}

\begin{figure}[H]
    \centering
    \includegraphics[scale=0.50]{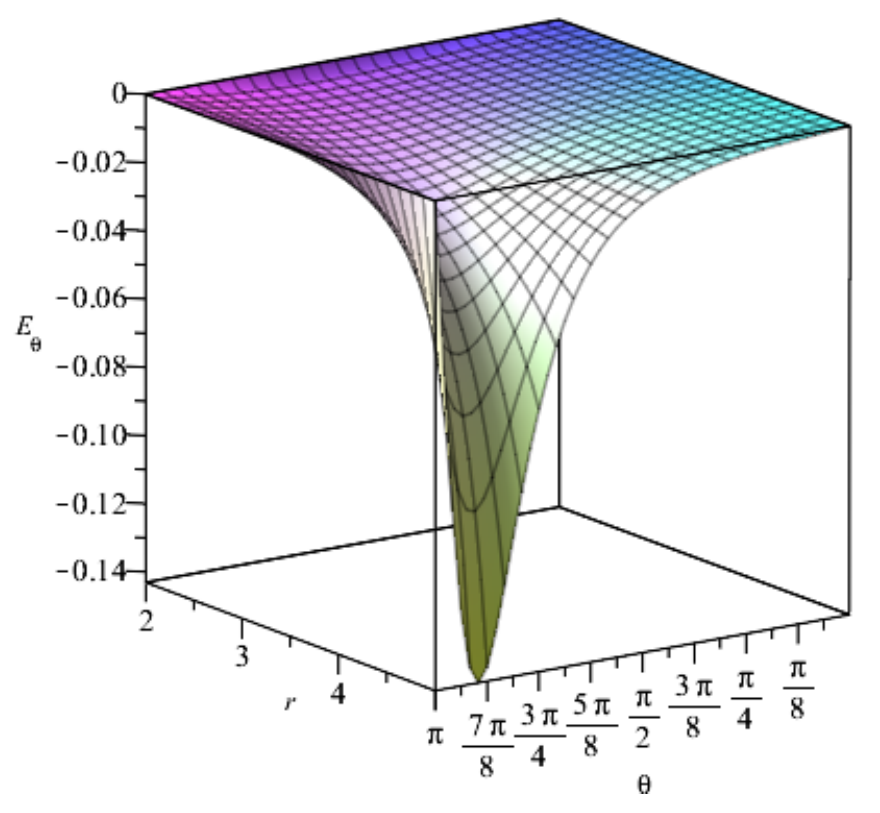}
    \caption{The $\theta$-component of  the electric field  for $t=1$, as a function of the coordinates $r$ and $\theta$, where we set the constants as $j_+=0.5$, $j_-=15$, $a=1$, $\eta=1$, $\vartheta=2$, $c=1$.}
    \label{fig:etab}
\end{figure}

According to the action (\ref{eq:actionab}), as the coupling constant $a$ increases, the strength of the interaction between the dilaton field and the electromagnetic field decreases, while the strength of the interaction between the dilaton and the cosmological constant increases. Moreover, based on equation (\ref{cosmoc}), the cosmological constant $\Lambda$ can take negative, positive or zero values depending on 
the coupling constant $a$. It is known that in asymptotically AdS/dS spacetimes,  the near boundary or the deep events are holographically dual to   the conformal field theory \cite{leblond2002tall}. We can interpret the holography in terms of renormalization group flows in the context of the  c-theorem in asymptotically dS spacetimes \cite{ghezelbash2010cosmological}. Based on the c-theorem, the renormalization group flows the ultraviolet for any expanding dS spacetime and to the infrared in any contracting dS spacetime \cite{strominger2001inflation, balasubramanian2002mass}. The c-function is given by  \cite{ghezelbash2015cosmological},
\begin{equation}
    c\sim \frac{1}{{G_{t t}}^{3/2}},
\end{equation}
We plot the c-function and show its behaviour with respect to the time coordinate, in order to infer the dS spacetime as an expanding or contracting. We present the c-function in figure \ref{fig:cab1} for the five-dimensional spacetime (\ref{st}), where the coupling constant $a>1$. We note that our five-dimensional spacetime extends by time. 
It is noteworthy that our solutions to the Einstein-Maxwell-dilaton theory based on the four-dimensional Bianchi type IX metric, are completely independent of the constant $k$ (which appears in the Bianchi type IX geometry and belongs to the interval $0\leq k\leq 1$). Moreover, we present the behaviour of the dilaton field with respect to the coordinates $r$ and $\theta$ for two different time slices in figure \ref{fig:phab}.
Accourding to our solutions, we show the components of the electric field in figures \ref{fig:erab} and \ref{fig:etab}, with respect to the coordinates $r$ and $\theta$.
\section{Exact solutions to the Einstein-Maxwell-dilaton based on the Bianchi type IX geometry with equal coupling constants $a$ and $b$}
We find and analyze the exact solutions to the five-dimensional Einstein-Maxwell-dilaton theory based on the four-dimensional Bianchi type IX geometry. In this section, we consider the coupling constants $a$ and $b$ to be equal to each other. For this case, a new set of ansatzes for the five-dimensional metric, the electromagnetic field and the dilaton field is needed, as the considered ansatzes (\ref{st})-(\ref{dil}) leads to the following constraint on the coupling constants $ab=-2$, which cannot be satisfied for $a=b$. First, we consider the case where the coupling constants are equal to each other and non-zero, and then, we consider the case where the coupling constants are both zero. The second case leads to the solutions for the Einstein-Maxwell theory in the presence of the cosmological constant.
\subsection{Exact solutions where the non-zero coupling constants are equal}
We consider the following ansatzes for the five-dimensional metric, the electromagnetic field and the dilaton field as,
\begin{equation}
    ds_5^2=-\frac{1}{H^2(t,r,\theta)}dt^2+R^2(t)H(t,r,\theta)ds_{B. IX}^2, \label{ds}
\end{equation}
\begin{equation}
  A_t(t,r,\theta)=\alpha R^M(t)H^E(t,r,\theta)  , \label{At}
\end{equation}
\begin{equation}
    \phi(t,r,\theta)=-\frac{3}{4a}\ln{(H^L(t,r,\theta)R^W(t))}, \label{ph}
\end{equation}
where $ds^2_{B.IX}$ is the Bianchi type IX geometry given in (\ref{trbix2}) and $M$, $E$, $L$ and $W$ are constants. In these new set of ansatzes, the metric function $H(t,r,\theta)$ depends on time coordinate, as well as the spatial coordinates $r$ and $\theta$. 
We find the constants $M$ and $E$ through the Einstein and Maxwell field equations, as $M=-a^2$ and $E=\frac{-a^2}{2}-1$. Moreover, the constants $L$ and $W$ in dilaton field are found to be $L=a^2$ and $W=2a^2$. By analyzing the $\varepsilon_{r r}$ component of the Einstein field equations, we find the solutions to the metric function $R(t)$, as given by,
\begin{equation}
    R(t)=(\epsilon t + \mu)^{1/a^2}. \label{RRR}
\end{equation}
In equation (\ref{RRR}), $\epsilon$ and $\mu$ are arbitrary constants. Solving the other Einstein equations, we find the metric function $H(t,r,\theta)$ as,
\begin{equation}
    H(t,r,\theta)=(R^{a^2+2}(t)+G(r,\theta))^{\frac{2}{a^2+2}}R^{-2}(t), \label{HH}
\end{equation}
where $R(t)$ is given by equation (\ref{RRR}) and the metric function $G(r,\theta)$ has the following form,
\begin{equation}
    G(r,\theta)=g_+r^2\cos{\theta}+g_-,\label{GG}
\end{equation}
In equation (\ref{GG}), $g_+$ and $g_-$ are arbitrary constants.  Moreover, we find the cosmological constant $\Lambda$ as,
\begin{equation}
    \Lambda=\frac{3\epsilon^2(4-a^2)}{2a^4}. \label{cosmo}
\end{equation}

\begin{figure}[H]
    \centering
    \includegraphics[scale=0.50]{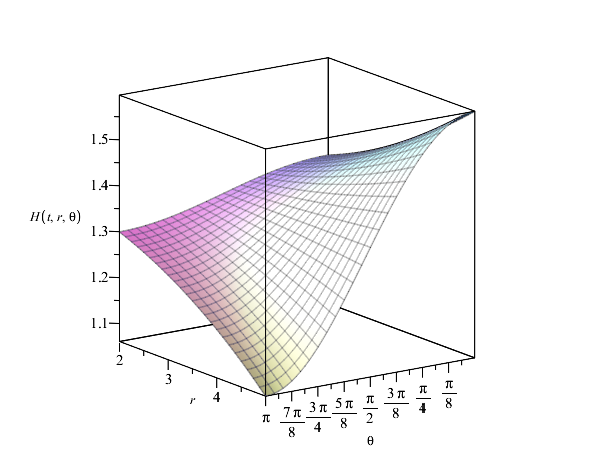}
    \caption{The behaviour of the metric function $H(t,r,\theta)$ with respect to the coordinates $r$ and $\theta$, where we consider specific values for the constants $\epsilon=1$, $\mu=2$, $g_+=0.5$, $g_-=15$ and $a=1$. }
    \label{figure:habe}
\end{figure}

\begin{figure}[htp]
    \centering
    \includegraphics[scale=0.50]{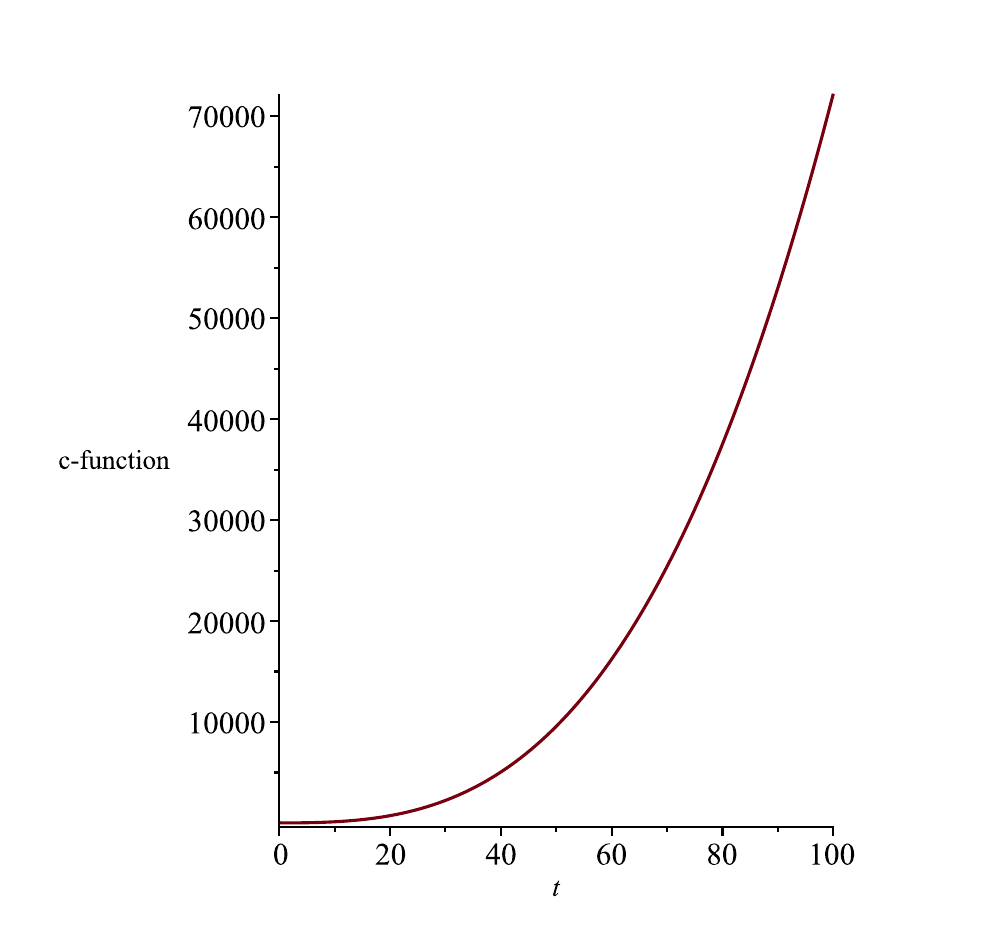}
  \caption{The behaviour of the c-function for the five-dimensional spacetime, where the coupling constants are equal to each other. }
    \label{figure:f1}
\end{figure}

\begin{figure}[htp]
    \centering
    \includegraphics[scale=0.50]{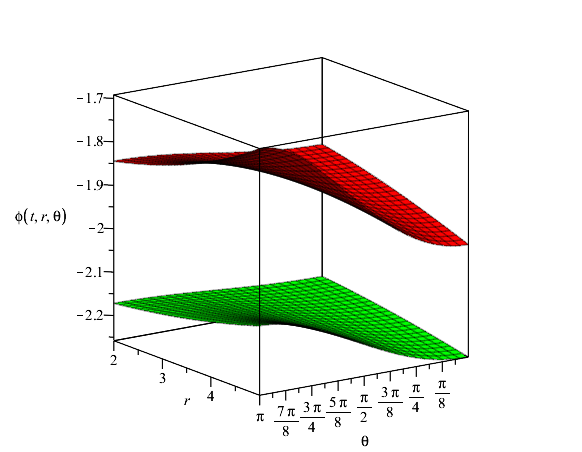}
     \caption{The dilaton field $\phi(t,r,\theta)$ for two different time slices $t=1$ and $t=2$ (upper and lower hypersurfaces respectively), where we set the constants $g_+=0.5$, $g_-=15$, $a=1$, $\epsilon=1$ and $\mu=2$. }
    \label{fig:phabe}
\end{figure}

\begin{figure}[H]
    \centering
    \includegraphics[scale=0.50]{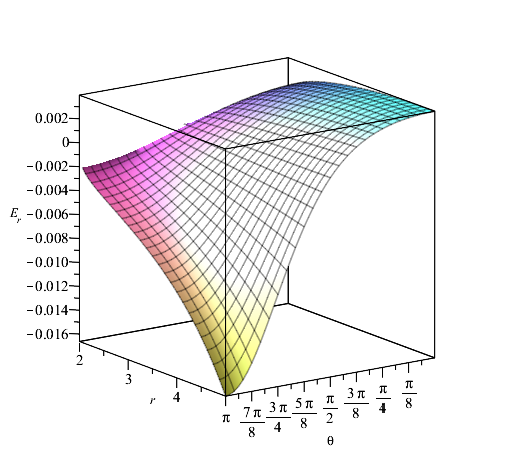}
    \caption{The behaviour of the $r$-component of the electric field with respect to the coordinates $r$ and $\theta$, for $t=1$. We consider the constants as $g_+=0.5$, $g_-=15$, $a=1$, $\epsilon=1$, $\mu=2$, $c=1$.}
    \label{fig:erabe}
\end{figure}

\begin{figure}[H]
    \centering
    \includegraphics[scale=0.50]{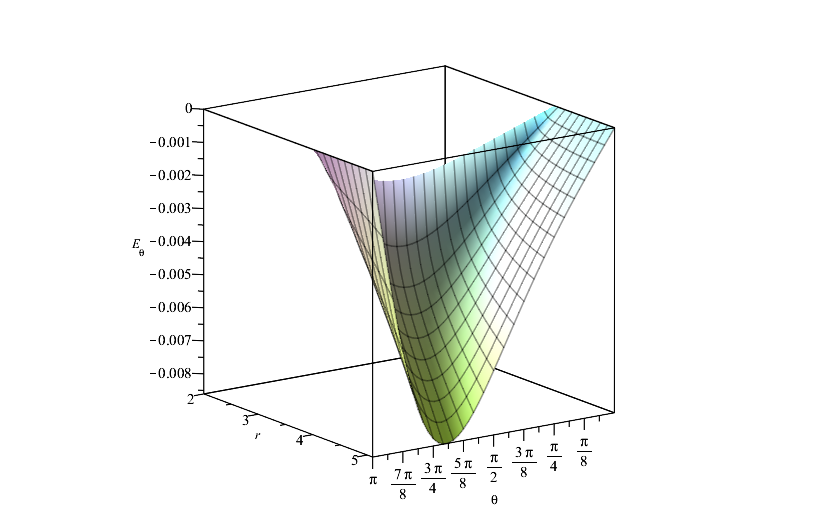}
    \caption{The behaviour of the $\theta$-component of the electric field with respect to the coordinates $r$ and $\theta$, where $t=1$. We consider specific values for the constants $g_+=0.5$, $g_-=15$, $a=1$, $\epsilon=1$, $\mu=2$, $c=1$.}
    \label{fig:etabe}
\end{figure}

According to the equation (\ref{cosmo}), the cosmological constant can take positive, negative or zero values based on the coupling constant $a$.We verify that all the components of the Einstein, Maxwell and dilaton field equations, satisfy with our solutions (\ref{RRR})-(\ref{cosmo}). The behaviour of the metric function $H(t,r,\theta)$ as a function of coordinates $r$ and $\theta$ is shown in figure \ref{figure:habe}. In this figure, we set the constants as $\epsilon=1$, $\mu=2$, $g_+=0.5$, $g_-=15$ and $a=1$.
 We present the changes of the c-function with respect to the time coordinate, in figure \ref{figure:f1}, where we considered the cosmological constant to be positive. We infer that the five-dimensional spacetime (\ref{ds}) expands in time for $t=$ constant slices. We note that the cosmological constant (\ref{cosmo}) and the dilaton field (\ref{ph}) diverge when the coupling constants $a$ and $b$ are equal to zero. Therefore, we need another way to find the exact solutions to the Einstein-Maxwell-dilaton theory, where the two coupling constants are equal to zero. Figure \ref{fig:phabe} indicates the behaviour of the dilaton field $\phi(t,r,\theta)$ for two different time slices. Moreover, we show the behaviour of the $r$ and $\theta$ components of the electric field in Figures \ref{fig:erabe} and \ref{fig:etabe}, respectively.


\subsection{Exact solutions where the coupling constant are equal to zero}
By considering the coupling constants $a$ and $b$ to be equal to zero, the dilaton field in the action (\ref{action5}) would decouple from the Einstein-Maxwell-dilaton theory, and the theory reduces to the Einstein-Maxwell theory with a cosmological constant. As we noted, since the dilaton field (\ref{ph}) and the cosmological constant (\ref{cosmo}) diverge in the limit of $a \rightarrow 0$, the exact solutions to the theory, for this case, cannot be obtained as a limit of the previous case, where $a=b \neq 0$. Therefore, we consider an ansatz for the electromagnetic gauge field as,
\begin{equation}
    A_t(t,r,\theta)=\frac{\alpha}{H(t,r,\theta)},\label{Atabzero}
\end{equation}
where $\alpha$ is an arbitrary constant. We also consider the same ansatz for the five-dimensional metric as in equation (\ref{ds}).

Through the Einstein field equations, we find the metric function $R(t)$, is given by,
\begin{equation}
    R(t)=\nu e^{\gamma R_0t},\label{Rabzero}
\end{equation}
where $\gamma=\pm1$, $R_0^2=\Lambda/6$ and $\nu$ is a constant. Also, from the $\epsilon_{t t}$ component of the Einstein equation, the constant $\alpha$ in the electromagnetic ansatz is found to be $\alpha=(3/2)^{1/2}$. Moreover, we find the metric function $H(t,r,\theta)$,
\begin{equation}
  H(t,r,\theta)=1+(f_+r^2\cos{\theta}+f_-)e^{\frac{-\gamma \sqrt{6\Lambda}t}{3}}. \label{HHH}
\end{equation}
In equation (\ref{HHH}), $f _\pm$ are arbitrary constants. We verify explicitly that all the other Einstein and Maxwell equations are satisfied.

\begin{figure}[H]
    \centering
    \includegraphics[scale=0.50]{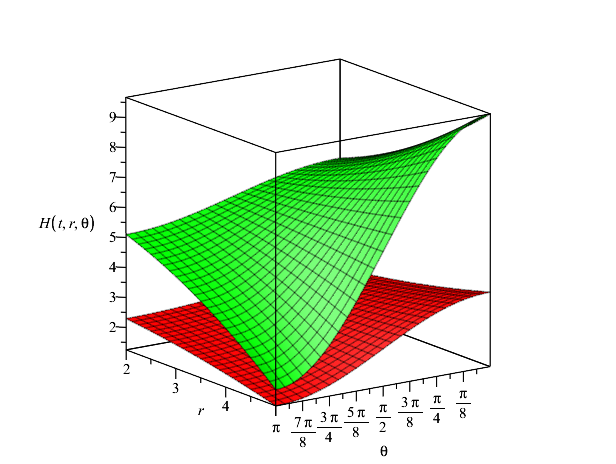}
    \caption{The metric function $H(t,r,\theta)$ as a function of the spatial coordinates $r$ and $\theta$ for two different time slices $t=1$ and $t=2$ (upper and lower surfaces, respectively), where we consider specific values for the constants $f_+=0.5$, $f_-=15$ and $\Lambda=2$ and we consider $\gamma=+1$.}
    \label{fig:habez}
\end{figure}

\begin{figure}[htbp]
    \centering
    \includegraphics[scale=0.50]{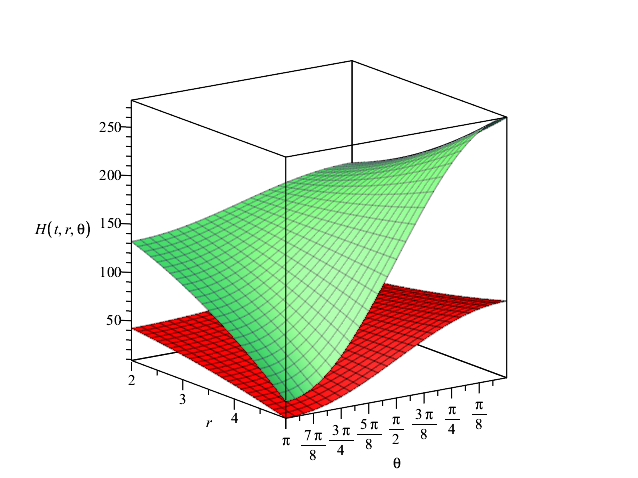}
    \caption{The behaviour of the metric function $H(t,r,\theta)$ as a function of the coordinates$r$ and $\theta$ for two different time slices $t=1$ and $t=2$ (lower and upper surfaces, respectively), where $\gamma=-1$ and we set the constants as $f_+=0.5$, $f_-=15$ and $\Lambda=2$.}
    \label{fig:habez2}
\end{figure}

We represent the behaviour of the metric function $H(t,r,\theta)$ in figures \ref{fig:habez} and \ref{fig:habez2}, with respect to the coordinates $r$ and $\theta$, where we set the constants $\gamma=+1$ and $\gamma=-1$, respectively. According to the figures, the metric function decreases monotonically with time where $\gamma=+1$, and increases monotonically where $\gamma=-1$.

\begin{figure}[htb!]
\centering
\includegraphics[scale=0.50]{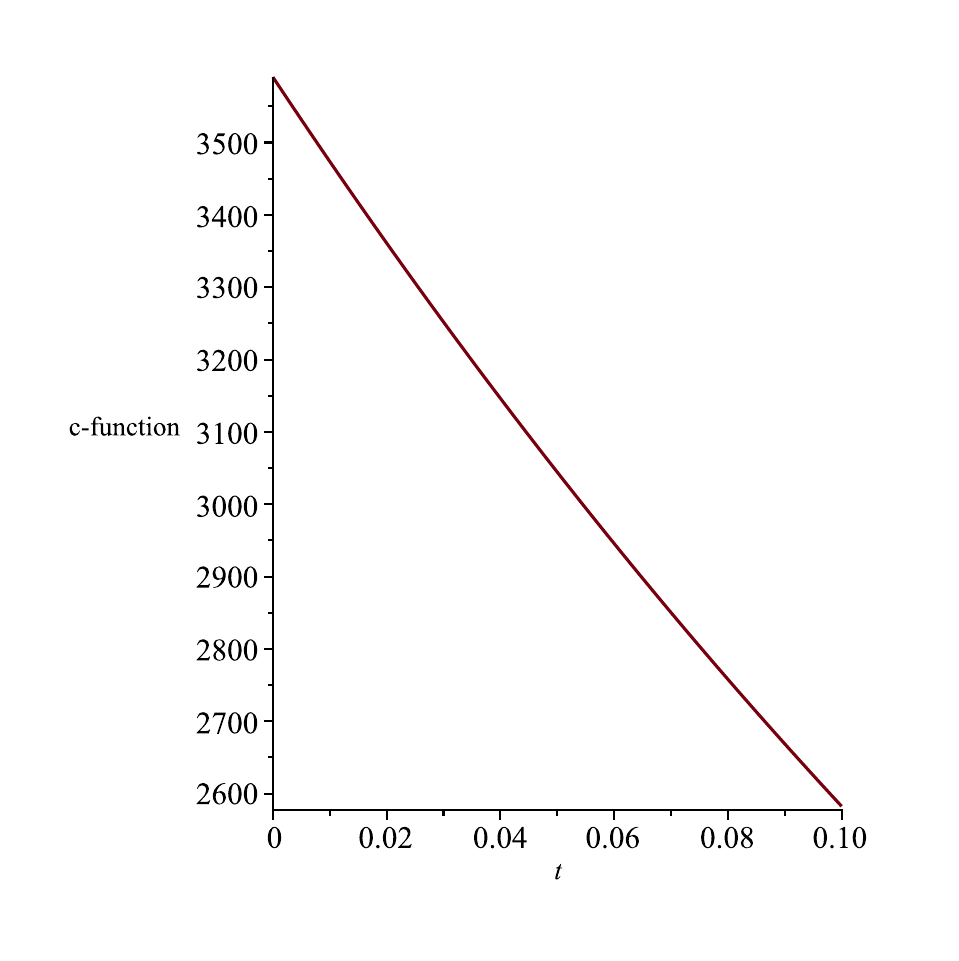}
\includegraphics[scale=0.50]{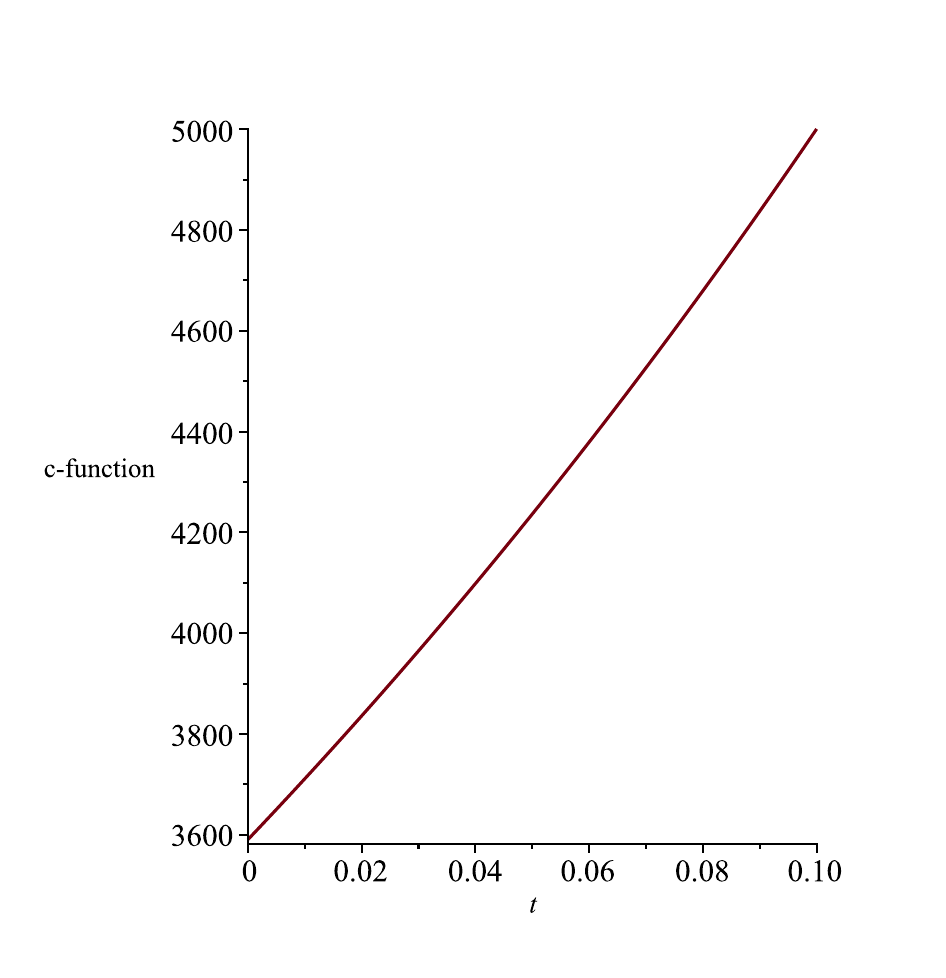}
\caption{The c-functions with $\gamma=+1$ (left) and $\gamma=-1$ (right). }
\label{fig:cabz}
\end{figure}

\begin{figure}[htbp]
    \centering
    \includegraphics[scale=0.50]{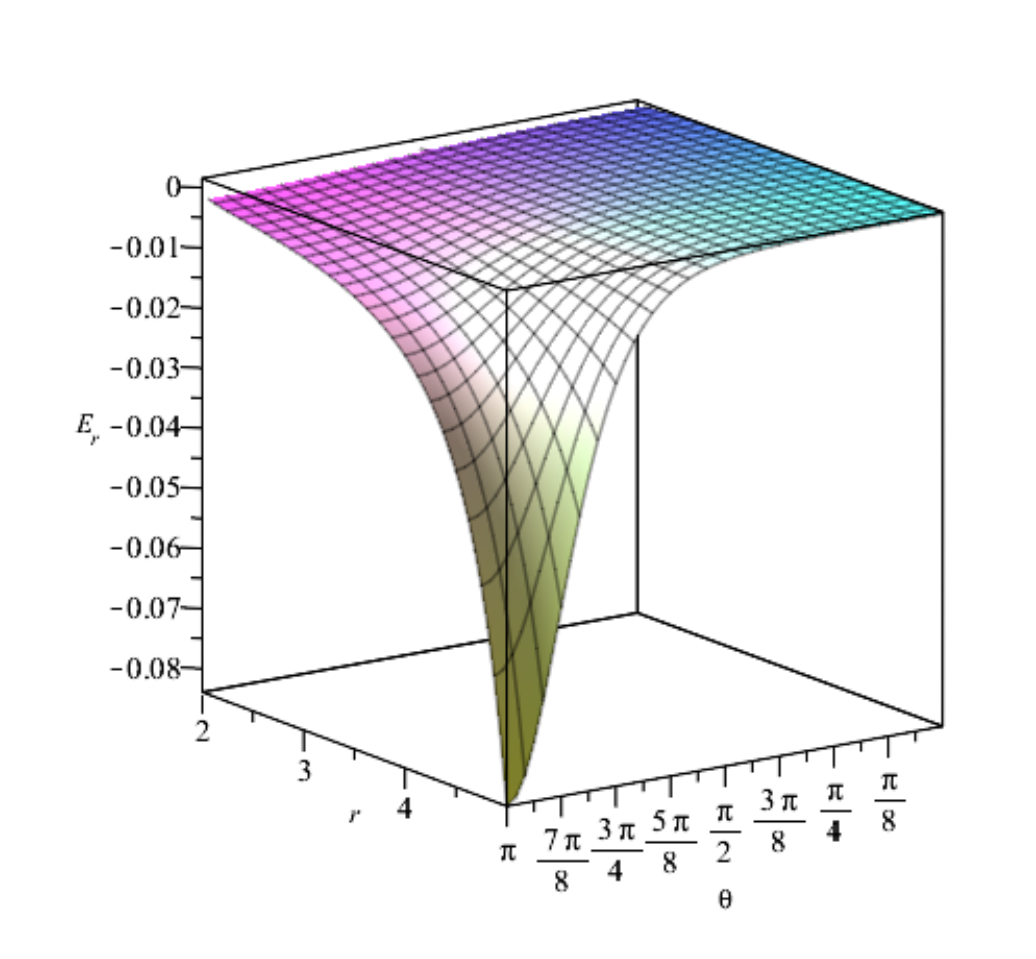}
    \caption{The behaviour of the $r$-component of electric field with respect to $r$ and $\theta$ coordinates for $t=1$. The constants are set as $f_+=0.5$, $f_-=15$, $\Lambda=2$, $\nu=3$ and $c=1$.}
    \label{fig:erabez}
\end{figure}

\begin{figure}[htbp]
    \centering
    \includegraphics[scale=0.50]{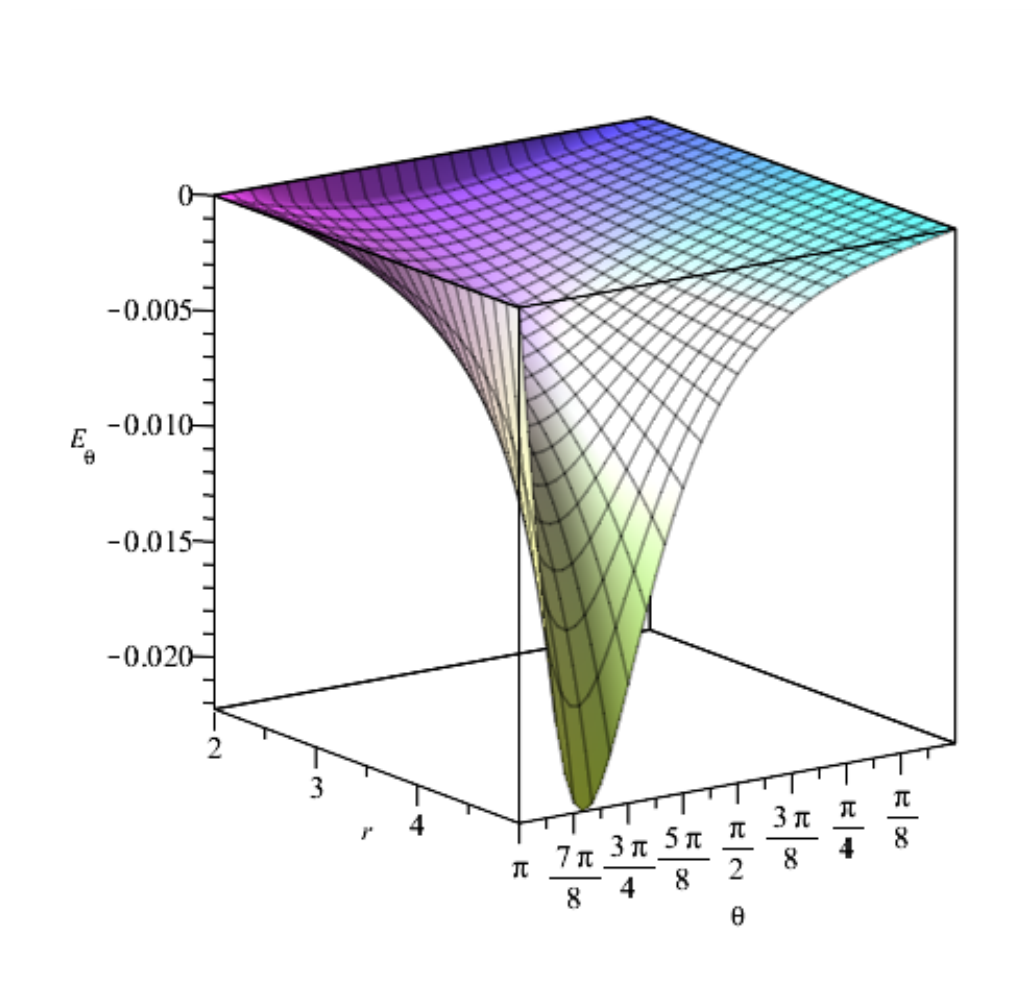}
    \caption{The behaviour of the $\theta$-component of electric field with respect to $r$ and $\theta$ coordinates for $t=1$. The constants are set as $f_+=0.5$, $f_-=15$, $\Lambda=2$, $\nu=3$ and $c=1$.}
    \label{fig:etabez}
\end{figure}

By presenting the c-function in figure \ref{fig:cabz} for two different values of $\gamma=+1$ and $\gamma=-1$, we notice that the $t=$ constant slices of the spacetime are expanding where $\gamma=-1$, and contracting where $\gamma=+1$. The behaviour of the $r$ and $\theta$ components of the electric field, is presented in figures \ref{fig:erabez} and \ref{fig:etabez}, respectively.
\section{More general solutions for the Einstein-Maxwell-dilaton theory based on Eguchi-Hanson type II geometry}
The exact solutions to the Einstein-Maxwell-dilaton theory (\ref{action5}) based on the four-dimensional Eguchi-Hanson type II geometry are well-know \cite{ghezelbash2017new}. The Eguchi-Hanson type II geometry is an important subspace of the Bianchi type IX geometry. Hence, we present a more general class of solutions to the Einstein-Maxwell-dilaton theory based on this geometry, for three different cases, where the coupling constants $a$ and $b$ are non-zero and not equal to each other, where $a=b \neq 0$ and where $a=b=0$.
\subsection{More general solutions where  the coupling constants are {\color{black}  not} equal  }
We consider an ansatz for the five-dimensional spacetime as \cite{ghezelbash2017new,mahapatra1999eguchi},
\begin{equation}
    ds_5^2=-\frac{1}{H_{EH}^2(r,\theta)}dt^2+R^2(t)H_{EH}(r,\theta)ds_{EH.II}^2,
\end{equation}
where $ds_{EH.II}^2$ represents the Eguchi-Hanson type II geometry (given in equation (\ref{EHIIBB})). We consider the same ansatzes for the electromagnetic gauge field and the dilaton field, as given in
 equations (\ref{A}) and (\ref{dil}), respectively. The solutions for the metric function $H_{EH}(r,\theta)$ is given by \cite{ghezelbash2017new},
\begin{equation}
      H_{EH}(r,\theta)=(1+\frac{g_+}{r^2+h^2\cos{\theta}}+\frac{g_-}{r^2-h^2\cos{\theta}})^{\frac{2}{2+a^2}},\label{HEH}
\end{equation}
where $g_+$ and $g_-$ are arbitrary constants. We note that the solution (\ref{HEH}) for the metric function, is not a solution for the Einstein-Maxwell-dilaton theory based on the Bianchi type IX space, unless the constant $k$ in the Bianchi metric (\ref{trbix2}) is equal to one. It is worth noting that as the field equations are nonlinear, the linear summation of  (\ref{H}) and (\ref{HEH}), is not a solution to the theory. 

We find a general solution to the Einstein-Maxwell-dilaton theory, where the dilaton field is coupled to both the electromagnetic field and the cosmological constant, based on the four-dimensional Eguchi-Hanson type II space, which is given by the metric function as,
\begin{equation}
  {\cal H}_{EH}(r,\theta)=(j_+r^2\cos{\theta}+j_-+\frac{g_+}{r^2+4c^2\cos{\theta}}+\frac{g_-}{r^2-4c^2\cos{\theta}})^{\frac{2}{2+a^2}},
\end{equation}
where $a$ is the coupling constant and $g_\pm$, $j_\pm$ and $c$ are arbitrary constants. The metric function $R(t)$ and the cosmological constant are still given by (\ref{RR}) and (\ref{cosmoc}). Moreover, we get the same constraint on the coupling constant $a$ and $b$, as given by equation (\ref{abm2}).
\subsection{More general solutions where the non-zero coupling constants are equal}
We consider the following ansatz for the five-dimensional metric,
\begin{equation}
     ds_5^2=-\frac{1}{H_{EH}^2(t,r,\theta)}dt^2+R^2(t)H_{EH}(t,r,\theta)ds_{EH.II}^2, \label{ehh}
 \end{equation}
where the two coupling constants are non-zero and equal to each other. We consider the same ansatzes for the electromagnetic field and the dilaton field, as in equations (\ref{At}) and (\ref{ph}), respectively. 
 
 The metric function $H_{EH}(t,r,\theta)$ is found in  \cite{ghezelbash2017new}, and is given by,
 \begin{equation}
      H_{EH}(t,r,\theta)=(R^{a^2+2}(t)+K(r,\theta))^{\frac{2}{a^2+2}}R^{-2}(t), \label{HEH2}
 \end{equation}
where the function $R(t)$ is given by,
\begin{equation}
      R(t)=(\epsilon t + \mu)^{1/a^2},\label{R111}
\end{equation}
and the function $K(r,\theta)$ is,
\begin{equation}
     K(r,\theta)=1+\frac{k_+}{r^2+h^2\cos{\theta}}+\frac{k_-}{r^2-h^2\cos{\theta}}. \label{k}
\end{equation} 
In equations (\ref{R111}) and (\ref{k}), $\epsilon$, $\mu$ and $k_\pm$ are arbitrary constants \cite{ghezelbash2017new}. We note that the linear summation of  (\ref{HH}) and (\ref{HEH2}) 
is not a general solution to the theory.

We find a combined solution for the metric function for the Einstein-Maxwell-dilaton theory, based on the four-dimensional Eguchi-Hanson type II geometry, as given by,
\begin{equation}
   {\cal  H}(t,r,\theta)={\frac {1}{ \left( R \left( t \right)  \right) ^{2}} \left(  \left( R
 \left( t \right)  \right) ^{{a}^{2}+2}+K \left( r,\theta \right) +{
\it G(r,\theta)} \right) ^{\,  \frac{2}{2+a^2} }},
\end{equation}
where  $G(r,\theta)$ is given in equation (\ref{GG}). We also find that the cosmological constant, is still given by equation (\ref{cosmo}), which can take positive, negative or zero values, based on the coupling constant.
\subsection{More general solutions where the coupling constants are zero
}
We consider ansatzes for the five-dimensional metric and the electromagnetic field as in (\ref{ehh}) and (\ref{Atabzero}), respectively. The metric functions $H_{EH}(t,r,\theta)$ is found to be as \cite{ghezelbash2017new},
\begin{equation}
    H_{EH}(t,r,\theta)=1+\exp({\frac{-\gamma\sqrt{6\Lambda}t}{3}})\{\frac{a_+}{r^2+4c^2\cos{\theta}}+\frac{a_-}{r^2-4c^2\cos{\theta}}\},\label{HEH3}
    \end{equation}
where $a_\pm$ are two constants and $\gamma=\pm1$. The metric function $R(t)$ is given the same as (\ref{Rabzero}). We find a general solution for the Einstein-Maxwell-dilaton theory based on the Eguchi-Hanson type II space, given by the metric function ${\cal H}(t,r,\theta)$ as,
\begin{equation}
       {\cal H}(t,r,\theta)=1+\exp({\frac{-\gamma\sqrt{6\Lambda}t}{3}})\{\frac{a_+}{r^2+4c^2\cos{\theta}}+\frac{a_-}{r^2-4c^2\cos{\theta}}+j_+r^2\cos{\theta}+j_-\},
\end{equation}
where $j_+$ and $j_-$ are constants. It is worth noting that by choosing $\gamma=+1$, the metric function ${\cal H}(t,r,\theta)$ decreases monotonically in time, while choosing $\gamma=-1$ makes the metric function ${\cal H}(t,r,\theta)$ increases monotonically in time.

\section{Uplifting to higher dimensions}
In this section, we study three different uplifting process for the solutions to the five-dimensional Einstein-Maxwell-dilaton theory.

First, we consider the uplifting of the solutions to higher than five-dimensional theories, for the case where the coupling constants are not equal.
Uplifting the solutions of the five-dimensional Einstein-Maxwell-dilaton theory to the Einstein-Maxwell theory in higher than five dimensions ($5 + {\cal D}$ dimensions) is possible only if ${\cal D}$ satisfies the following equation
\cite{ghezelbash2017new,gouteraux2012holography,gouteraux2011generalized},
\begin{equation}
 {\cal D}=\frac{3a^2}{1-a^2}.\label{DD}
\end{equation}
Moreover, the uplifting is only possible if the coupling constants $a$ and $b$ are equal.  The latter condition violates the constrain that we found for the coupling constants (\ref{abm2}). Therefore, the solutions to the Einstein-Maxwell-dilaton theory cannot be uplifted to higher than five-dimensional Einstein-Maxwell theory.

The other uplifting process is to uplift the solutions to the five-dimensional Einstein-Maxwell-dilaton theory to the solutions of the six-dimensional Einstein gravity in the presence of the cosmological constant. 
In order to uplift the solutions in this case, the coupling constants have to be $a=\pm 2$ and $b=\pm \frac{1}{2}$ \cite{charmousis2009einstein}. Hence, the coupling constants satisfy $ab=1$, which is in contrast to the equation (\ref{abm2}). 

We conclude that the solutions to the Einstein-Maxwell-dilaton theory where the coupling constants are not equal, cannot be uplifted to those of the Einstein-Maxwell theory or the Einstein gravity in higher dimensions.

We analyze another approach for the uplifting process, based on the reference \cite{gouteraux2011generalized}. We consider the action for the Einstein gravity in presence of a cosmological constant $\Lambda_D$ in $D$-dimensions as \cite{ghezelbash2017new},
\begin{equation}
    S_D=\int d^Dx\sqrt{-g}(R-\frac{1}{2(q+2)!}\mathcal{F}^2_{[q+2]}+2\Lambda_D), \label{uac}
\end{equation}
where $D=p+q+1$ and $\mathcal{B}_{[q+1]}$ is a $q+1$-potential. Moreover, in equation (\ref{uac}), $R$ represents the Ricci scalar for the $D$-dimensional spacetime and $\mathcal{F}_{[q+2]}$ is given as,
\begin{equation}
    \mathcal{F}_{[q+2]}=d\mathcal{B}_{[q+1]},
\end{equation}
where $\mathcal{F}_{[q+2]}$ is a $q+2$-field strength form and $d\mathcal{B}_{[q+1]}$ is the exterior derivation of the $\mathcal{B}_{[q+1]}$ potential.

We consider the dimensional reduction from $D$-dimensions to $p+1$-dimensions on an internal curved $q$-dimensional space, where we show the line element of the internal $q$-dimensional space by $d\mathcal{K}^2_q$ \cite{ghezelbash2017new}. By considering the following ansatzes for the $D$-dimensional metric and the $q+1$-potential $\mathcal{B}_{[q+1]}$,
\begin{equation}
    ds^2_D=e^{-\delta\phi'}ds^2_{p+1}+e^{\phi'(\frac{2}{\delta(p-1)}-\delta)}d\mathcal{K}^2_q, \label{ddim}
\end{equation}
\begin{equation}
    \mathcal{B}_{[q+1]}=\mathcal{A}_{[1]}\wedge d\mathcal{K}_q. \label{qpo}
\end{equation}
We consider the Einstein-Maxwell-dilaton theory in $p+1$ dimensions with a potential as \cite{ghezelbash2017new},
\begin{equation}
    S_{p+1}=\int d^{p+1}x(R'-\frac{1}{2}(\nabla\phi')^2-\frac{1}{4}e^{\gamma\phi'}\mathcal{F}^2_{[2]}+2\Lambda_De^{-\delta\phi'}+2\Lambda'e^{-\frac{2}{\delta(p-1)}\phi'}). \label{pdimon}
\end{equation}
In equation (\ref{pdimon}), $R'$ represents the Ricci scalar for the $p+1$-dimensional spacetime and $\Lambda'=R''/2$, where $R''$ is the Ricci scalar of the internal space. In action (\ref{pdimon}), $\delta$ and $\gamma$ are the dilaton coupling constants, which have the following relations \cite{gouteraux2011generalized},
\begin{equation}
    \delta=(\frac{2q}{(p-1)(p+q-1)})^{1/2},
\end{equation}
\begin{equation}
    \gamma=\delta(2-p).
\end{equation}
By comparing the action in equation (\ref{pdimon}) and (\ref{action5}) and redefining the dilaton field and considering the following relations for the coupling constants, we note that our solutions for the five-dimensional metric, electromagnetic field and dilaton field can be uplifted to a higher dimensional theory in the absence of the cosmological constant $\Lambda_D=0$,
\begin{equation}
    \phi'=2\sqrt{\frac{2}{3}}\phi,
\end{equation}
\begin{equation}
    \delta=-\frac{3}{4}\sqrt{\frac{3}{2}}\frac{1}{(p-1)b},
\end{equation}
\begin{equation}
    \gamma=-\frac{2}{3}\sqrt{\frac{3}{2}}a.
\end{equation}
By considering the constraint that we found on the coupling constants $ab=-2$, we find that $p=4$. Moreover, for having the exact same action as (\ref{action5}), we consider,
\begin{equation}
    \mathcal{A}_{[1]}=2A_tdt,
\end{equation}
\begin{equation}
    2\Lambda'=-\Lambda.
\end{equation}
Therefore, we can uplift our solutions to a higher dimensional theory without a cosmological constant.

Now we study the uplifting of the solutions to the five-dimensional Einstein-Maxwell-dilaton theory to higher than five-dimensional theories, where the non-zero coupling constants are equal.The solutions to the five-dimensional Einstein-Maxwell-dilaton theory to higher $(5+{\cal D})$ dimensional Einstein-Maxwell theory with a cosmological constant, is possible only if equation (\ref{DD}) holds. In order to have ${\cal D} \geq 1$, we find the coupling constant $a$ satisfies,
\begin{equation}
    \frac{1}{2}\leq a <1. \label{coupling}
\end{equation}
The range of coupling constant $a$ as given by (\ref{coupling}), is in contrast to the condition (\ref{constraint}) for the coupling constant. 
Therefore, the uplifting of solutions to the Einstein-Maxwell-dilaton theory to a higher dimensional Einstein-Maxwell theory with a cosmological constant is not possible.
\section{Conclusions}
We found a class of exact solutions to the five-dimensional Einstein-Maxwell-dilaton theory based on the Bianchi type IX geometry, where the dilaton field is coupled to both the electromagnetic field and the cosmological constant, with two different coupling constants. We considered ansatzes for the five-dimensional metric, electromagnetic field and the dilaton field. We solved all the field equations and determined the metric functions. Through the field equations, we found a relation between the coupling constants. Moreover, we obtained an extra constraint on the coupling constant, which indicates that it only 
takes some specific values. 
By calculating the cosmological constant, we noted that it can be positive, zero or negative. We studied the c-function for the five-dimensional metric. 
By calculating the Kretschmann invariant, we discussed the singularities of the spacetime. 

Moreover, we showed that the solutions cannot be uplifted to a higher dimensional Einstein gravity or a higher dimensional Einstein-Maxwell theory in the presence of a cosmological constant. We also found another class of the exact solutions to the Einstein-Maxwell-dilaton theory where the coupling constants are equal. We then considered the limit where both the coupling constants are zero and found a new class of exact solutions to the theory.
Based on the different classes of the exact solutions, we found a more general class of solutions to the Einstein-Maxwell-dilaton theory on the transverse Eguchi-Hanson type II geometry.
{\color{black}We also mention that the class of exact solutions, in this article, is unique and exists only in five dimensional Einstein-Maxwell-dilaton theory. Extending the solutions to six and higher dimensions ($N \geq 5$), by including an Euclidean space to the existing Bianchi type IX geometry (such as equation (\ref{st6}) in $N=5$) lead to trivial solutions for the metric functions $H(r,\theta)$ and $R(t)$. One remedy to find the higher dimensional solutions to the Einstein-Maxwell-dilaton theory might be considering additional fields, such as extra vector fields to the standard Einstein-Maxwell-dilaton theory \cite{add2}. The added fields may support the existence of the solutions, in dimensions greater than five.} 

 {\color{black}We conclude with the observation that the well-known holography between the rotating black holes and the conformal field theories (CFTs) enjoys the independence of the central charges of the CFT on the non-gravitational matter fields \cite{Add3, Add4}. In this article, we found some exact solutions to the five-dimensional gravity coupled to non-gravitational fields. Using the Janis–Newman algorithm, we can find the rotating versions of the exact solutions, presented in the article. The rotating solutions provide a treasure trove of solutions, including black holes, where we can find and study their holographic dual CFTs. Moreover, we can test the independence of the central charges of the CFT on the non-gravitational fields, for a broader class of gravitational theories.  One other interesting line of research is to seek the possible hidden symmetry in the solutions space of a probe field, in the background of rotating versions of the exact solutions, presented in the article. These symmetries, in general, lead to finding the possible dual hidden CFT to the black holes \cite{Add5}. Moreover extending the dual hidden CFT by introducing a deformation parameter in the radial equation of the probe field, as well as finding the different pictures for the dual hidden CFT \cite{Add6}, are some of other applications of the rotating versions of the exact solutions, presented in the article.
We leave studying the rotating versions of the exact solutions and their above-mentioned applications in holography for a future article.  } 
 {\color{black}
 \begin{acknowledgments}
This work is supported by the Natural Sciences and Engineering Research Council of Canada. 
\end{acknowledgments} 
}
\begin{appendix}
 {\color{black}
\section{The Einstein-Maxwell-dilaton theory in six and higher dimensions ($N \geq 5$)}
In this appendix, we first consider the Einstein-Maxwell-dilaton theory (\ref{action}) in six dimensions, where $N=5$. We consider an ansatz for the six-dimensional metric as,
\begin{equation}
    ds_6^2=-\frac{1}{H^2(r,\theta)}dt^2+R^2(t)H(r,\theta)(ds_{B. IX}^2+dy^2). \label{st6}
\end{equation}
In the ansatz (\ref{st6}), $ds_{B.IX}^2$ represents the four-dimensional Bianchi type IX metric given by equation (\ref{trbix2}), and $H(r,\theta)$ and $R(t)$ are two metric functions, and $y$ is the sixth coordinate.

We consider the electromagnetic gauge field and the dilaton field in terms of the metric functions $H(r,\theta)$ and $R(t)$ as,
\begin{equation}
  A_t(t,r,\theta)=\alpha_6 R^{M_6}(t)H^{E_6}(r,\theta)  , \label{A6}
\end{equation}
\begin{equation}
    \phi(t,r,\theta)=-\frac{1}{a}\ln{(H^{L_6}(r,\theta)R^{W_6}(t))}, \label{dil6}
\end{equation}
where $\alpha_6$, $M_6$, $E_6$, $L_6$ and $W_6$ are arbitrary constants. According to the considered ansatz (\ref{A6}), the only non-zero component of the electromagnetic gauge field is the $t$ component, which is a function of time and spatial coordinates $r$ and $\theta$.
We find all the field equations (\ref{maxwell}), (\ref{dilaton}) and (\ref{ein}), where $N=5$. We get the following non-zero Einstein equations,
\begin{equation}
\epsilon_{r \phi}=16\,{\frac {{c}^{4}\sin \left( \psi \right) \cos \left( \psi \right) 
\sin \left( \theta \right) {r}^{3} \left( {\frac {\partial }{\partial 
\theta}}H \left( r,\theta \right)  \right)  \left( {k}^{4}-1 \right) 
}{H \left( r,\theta \right)  \left( 256\,{k}^{4}{c}^{8}-16\,{k}^{4}{c}
^{4}{r}^{4}-16\,{c}^{4}{r}^{4}+{r}^{8} \right) }},\label{e1}
\end{equation}
\begin{equation}
\epsilon_{\psi \theta}=-4\,{\frac {z(r,\psi) 
{c}^{4}\sin \left( \psi \right) 
\cos \left( \psi \right) {\frac {\partial }{\partial \theta}}H \left( 
r,\theta \right) }{{r}^{4}H \left( r,\theta \right)  \left( 16\,{c}^{4
}-{r}^{4} \right)  \left( 16\,{c}^{4}{k}^{4}-{r}^{4} \right) }},\label{e2}
\end{equation}
where,
\begin{eqnarray}
z(r,\psi) &=& 16( \cos^2 \psi )
{c}^{4}({k}^{8}-1){r}^{4}+
16 (\sin^2 \psi){c}^{4}({k}^{8}-1){r}^{4}+256\,{k}^{8}{c}^{8}-16\,{r}^{4}({k}^{8}-1){c}^{4}\nonumber\\
&-&
2( \cos ^2\psi)({k}^{4}-1){r}^{8}
-2\,
 (\sin^2 \psi)({k}^{4}-1){r}^{8}-256\,{k}^
{4}{c}^{8}+{r}^{8}({k}^{4}-1).\label{zzz}
\end{eqnarray}
Equations (\ref{e1}) and (\ref{e2}) imply that $H(r,\theta)=h(r)$. Plugging the latter equation for the metric function in the other Einstein equations $\epsilon_{r t}$, $\epsilon_{r r}$, $\epsilon_{t t}$ and $\epsilon_{\phi\phi}$, yields, 
\begin{equation}
h(r)=h_0,\, R(t)=R_0, \, \Lambda=0,
\end{equation}
where $h_0$ and $R_0$ are constants. So, we can not find any non-trivial metric function $H(r,\theta)$ for the six-dimensional Einstein-Maxwell-dilaton theory.
Comparing the Einstein equations for $N=4$ and $N=5$, we find the interesting point that, in the former theory the above-mentioned Einstein equations $\epsilon_{r \phi}$ and $\epsilon_{\psi \theta}$ are identically zero, while in the latter theory, they are not zero, where ultimately lead to trivial metric functions.

Moreover, in general, we consider the Einstein-Maxwell-dilaton theory (\ref{action}) in equal or higher than six dimensions, where $N \geq 5$. We consider an ansatz for the metric as,
\begin{equation}
    ds_N^2=-\frac{1}{H^2(r,\theta)}dt^2+R^2(t)H(r,\theta)(ds_{B. IX}^2+ds_{Euc.}^2), \label{st789}
\end{equation}
where $ds_{Euc.}^2$ is the ($N-4$)-dimensional Euclidean metric, with the coordinates $(y_1,\cdots ,y_{N-4})$.
A similar analysis of the field equations shows that the metric functions $H(r,\theta)$ and $R(t)$ can be only trivial, to satisfy all the field equations.

We also mention to find the higher dimensional solutions to the Einstein-Maxwell-dilaton theory, we might consider adding other fields, such as extra vector fields to the standard Einstein-Maxwell-dilaton theory \cite{add2}, to support the existence of the solutions, in dimensions greater than five. 

}
{\color{black}
\section{The Bianchi Classification of the homogeneous spaces}
In this appendix, following Bianchi's procedure, we represent the classification of the homogeneous Bianchi type spaces. We start with the group of transformations:
\begin{equation}
    x^\mu \rightarrow x'^\mu =T^\mu(x,\zeta),
\end{equation}
where the set of $\{\zeta^a|a\in 1,...,r\}$ are $r$ independent variables that parameterize the group. We consider an infinitesimal transformation:
\begin{equation}
      x^\mu \rightarrow x'^\mu =T^\mu(x,\zeta_0+\delta \zeta) \approx x^\mu+\xi^\mu_a(x)\delta\zeta^a=(1+\delta\zeta^a\xi_a)x^\mu, \label{trans}
\end{equation}
where $\zeta_0$ corresponds to the identical transformation, $T^\mu (x,\zeta_0)=x^\mu$. In equation (\ref{trans}) we consider \cite{imponente2002mixmaster}:
\begin{equation}
    (\frac{\partial T^\mu}{\partial\zeta^a})(x,\zeta_0)\equiv \xi_a^\mu (x).
\end{equation}

We define the $r$ first order differential operators $\{\xi_a\}$ in equation (\ref{trans}) in correspondence with the $r$ vectorial fields with components $\{\xi^\mu _a\}$ with the relation $\xi_a=\xi_a^\mu \frac{\partial}{\partial x^\mu}$, which are the killing generating vectors. The Lie algebra that applies to the killing vectors in the commutation relation form is given by:
\begin{equation}
    [\xi_a,\xi_b]=C^c_{ab}\xi_c \label{str},
\end{equation}
where $C^c_{ab}$ is the structure constant. By extending this formalism, we introduce the basis $\{e_\alpha\}$ for the Lie algebra  with the commutation relation as:
\begin{equation}
    [e_\alpha,e_\beta]=C^\gamma_{\alpha\beta}e_\gamma, \label{hom}
\end{equation}
and define the symmetric quantity as $\gamma_{\alpha \beta}=\gamma_{\beta \alpha}=C^\gamma_{\alpha\sigma}C^ \sigma_{\beta \gamma}$. The relation in equation (\ref{hom}) defines a group of transformation (non-Abelian) which represents the spatially homogeneous part of the spacetime. The metric tensor can be defined with respect to the bases $\{e_\alpha\}$ as:
\begin{equation}
    g_{\alpha\beta}=e_\alpha e_\beta,
\end{equation}
where $\{\alpha,\beta\in \{0,1,2,3\}\}$. Each class of the equivalence Lie group needs to be indicated by only one representative group. We show the tetradic basis of the four linearly independent vectors on each point as $e^i_{(a)}$, where $a\in \{1,...,4\}$ indicates the tetradic and $i$ represents the tensorial part. These bases satisfy the orthogonality condition:
\begin{equation}
    e_i^{(a)}e^k_{(a)}=\delta^k_i.
\end{equation}

Moreover, the metric tensor in terms of the tetradic bases is $g_{ij}=e_{(a)i}e_k^{(a)}=\eta_{(ab)}e_i^{(a)}e_k^{(b)}$. Hence, the line element becomes \cite{imponente2002mixmaster}:
\begin{equation}
    ds^2=\eta_{ab}(e^{(a)}_idx^i)(e_k^{(b)}dx^k).
\end{equation}
It is worth noting that $dx^{(a)}=e_i^{(a)}dx^i$ are not exact differentials of functions of the coordinates in general. The next step is to find the structure constants in a way that the metric becomes invariant under the homogeneity constraint. The Lie algebra in the tetradic bases is:
\begin{equation}
    [e_{a},e_{b}]=C^{c}_{a b}e_{c}, 
\end{equation}
In order to have an invariant line element $\gamma_{\alpha\beta}$ under the transformation of its group of motion, $\gamma_{\alpha\beta}$ must be the same under the homogeneity constraint. We write the spatial part of the line element as:
\begin{equation}
    dl^2=\eta_{ab}(e^{(a)}_\mu dx^\mu)(e^{(b)}_\nu dx^\nu), \label{lil}
\end{equation}
which makes $\gamma_{\mu \nu}$ to be $\gamma_{\mu \nu}=\eta_{ab}e^{(a)}_\mu e^{(b)}_\nu$. 
Equation (\ref{lil}) also implies that $e^{(a)}_\mu dx^\mu$ is invariant under such a transformation. Hence \cite{landau2013classical}:
\begin{equation}
    e^{(a)}_\mu(x) dx^\mu=e^{(a)}_\mu(x') dx'^\mu. \label{ex}
\end{equation}
A system of differential equations for determining $x'^\nu (x)$ can be obtained from (\ref{ex}) as:
\begin{equation}
    \frac{\partial x'^\nu}{\partial x^\mu}=e^\nu_{(a)}(x')e_\mu^{(a)}(x). \label{jj}
\end{equation}
The equation (\ref{jj}) is integrable if:
\begin{equation}
    \frac{\partial^2x'^\nu}{\partial x^\mu \partial x^\gamma}= \frac{\partial^2x'^\nu}{\partial x^\gamma \partial x^\mu}, \label{kk}
\end{equation}
which is called the Schwartz's condition \cite{minguzzi2015equality}. Substituting (\ref{jj}) in (\ref{kk}) we find:
\begin{equation}
    \frac{e^{(b)}_\gamma(x)e^{(a)}_\mu (x)}{e^\nu_{(a)}(x')}(\frac{\partial e_{(a)}^\nu (x')}{\partial x'^\sigma}e^\sigma_{(b)}(x')-\frac{\partial e_{(b)}^\nu (x')}{\partial x'^\sigma}e^\sigma_{(a)}(x'))=(\frac{\partial e^{(a)}_\gamma (x)}{\partial x^\mu}-\frac{\partial e^{(a)}_\mu (x)}{\partial x^\gamma}). \label{lon}
\end{equation}

After using the properties of the tetradic base and some algebra, we force both sides of the equation (\ref{lon}) to be equal to each other and equal to the same constant:
 \begin{equation}
     (\frac{\partial e_\mu^{(c)}}{\partial x^\nu}-\frac{\partial e_\nu^{(c)}}{\partial x^\mu})e^\mu _{(a)}e^\nu_{(b)}=C^c_{ab}, \label{sc}
 \end{equation}
where $C^c_{ab}$ is the structure constant. The uniformity condition is obtained by multiplying (\ref{sc}) by $e^\gamma_{(c)}$ \cite{imponente2002mixmaster}:
\begin{equation}
    e^\mu_{(a)}\frac{\partial e^\gamma_{(b)}}{\partial x^\mu}-e^\nu_{(a)}\frac{\partial e^\gamma_{(a)}}{\partial x^\nu}=C^c_{ab}e^\gamma_{(c)}. \label{imp}
\end{equation}
By defining a linear operator as $X_a=e^\mu_{(a)}\frac{\partial}{\partial x^\mu}$, we rewrite the equation (\ref{imp}):
\begin{equation}
    [X_a,X_b]=C^c_{ab}X_c,
\end{equation}
where the commutation $[X_a,X_b]$ implies $[X_a,X_b]=X_aX_b-X_bX_a$.
We use the Jacobi identity to express the homogeneity:
\begin{equation}
    [[X_a,X_b],X_c]+[[X_b,X_c],X_a]+[[X_c,X_a],X_b]=0. \label{jac}
\end{equation}

We can write equation (\ref{jac}) in terms of the structure constants:
\begin{equation}
(C_{ab}^hC^d_{ch}+C^h_{bc}C^d_{ah}+C^h_{ca}C^d_{bh})X_d=0,
\end{equation}
where the two index structure constant is defined as the dual of $C^c_{ab}$ as follow:
\begin{equation}
    C^c_{ab}=\epsilon_{abc}C^{dc}. \label{pss}
\end{equation}
In equation (\ref{pss}) $\epsilon$ is the Levi-Civita pseudo tensor and equation (\ref{jac}) can be written as:
\begin{equation}
\epsilon_{abc}C^{cd}C^{ba}=0.
\end{equation}
This mathematical tools and their relations enable us to classify the non-equivalent homogeneous spaces by using non-equivalent combinations of the constants $C^{ab}$ \cite{landau2013classical},
\begin{equation}
    [X_1,X_2]=-aX_2+n_3X_3, \label{comm2}
\end{equation}
\begin{equation}
    [X_2,X_3]=n_1X_1, \label{comm}
\end{equation}
\begin{equation}
    [X_3,X_1]=n_2X_2+aX_3. \label{comm3}
\end{equation}

The constants $(n_1,n_2,n_3)$ and $a$ in equations (\ref{comm2})-(\ref{comm3}) are related to the structure constants. Non-equivalent structure constants that lead to the non-equivalent homogeneous spaces is classified in Bianchi classification as follow \cite{landau2013classical}:

Type I: $a=0$ and $(n_1,n_2,n_3)=(0,0,0)$,

Type II: $a=0$ and $(n_1,n_2,n_3)=(1,0,0)$,

Type III: $a=1$ and $(n_1,n_2,n_3)=(0,1,-1)$,

Type IV: $a=1$ and $(n_1,n_2,n_3)=(0,0,1)$,

Type V: $a=1$ and $(n_1,n_2,n_3)=(0,0,0)$,

Type VI: $a=0$ and $(n_1,n_2,n_3)=(1,-1,0)$,

Type VII: $a=0$ and $(n_1,n_2,n_3)=(1,1,0)$,

Type VIII: $a=0$ and $(n_1,n_2,n_3)=(1,1,-1)$,

Type IX: $a=0$ and $(n_1,n_2,n_3)=(1,1,1)$.
 \vspace{0.5cm}

Each type of the Bianchi spaces has its own properties and applications in different theories. 
Among all of these Bianchi type spaces, we focus on the Bianchi type IX geometry, as it is the most symmetric space  (due to $(n_1,n_2,n_3)=(1,1,1)$), between all different types of the Bianchi geometry.
 
The Bianchi type IX geometry has essential properties and has been used in different areas of gravitational physics. This metric can be written with an $SO(3)$ or $SU(2)$ isometry group as \cite{ghezelbash2008supergravity}:
\begin{equation}
    ds^2_{B.IX}=e^{2f(\eta)}\sigma^2_1+e^{2h(\eta)}\sigma^2_2+e^{2g(\eta)}\sigma^2_3 + e^{2(f(\eta)+h(\eta)+g(\eta))}d\eta^2. \label{BIX}
\end{equation}
In equation (\ref{BIX}), $\sigma_i$'s are a basis of $SO(3)$ one-forms which satisfy $d\sigma_i=\frac{1}{2}\epsilon_{ijk}\sigma_j\sigma_k$ \cite{hanany2000anti}. Self-duality of the curvature gives first order differential equations for $f(\eta)$, $h(\eta)$ and $g(\eta)$  \cite{ghezelbash2008supergravity}:
\begin{equation}
    2\frac{df}{d\eta}=e^{2h}+e^{2g}-e^{2f}-2\lambda_1e^{h+g}, \label{l1}
\end{equation}
\begin{equation}
    2\frac{dh}{d\eta}=e^{2g}+e^{2f}-e^{2h}-2\lambda_2e^{f+g}, \label{l2}
\end{equation}
\begin{equation}
    2\frac{dg}{d\eta}=e^{2f}+e^{2h}-e^{2g}-2\lambda_3e^{f+h}. \label{l3}
\end{equation}
In equations (\ref{l1})-(\ref{l3}), the constants $  \{\lambda_i | i \in {1,2,3}\}$ satisfy the relation $\lambda_i \lambda_j=\epsilon_{ijk}\lambda_k$, which leads to five different choices for the set ($\lambda_1$,$\lambda_2$,$\lambda_3$) \cite{ghezelbash2006bianchi}. By choosing the constants as $(\lambda_1,\lambda_2,\lambda_3)=(0,0,0)$, the equations (\ref{l1}), (\ref{l2}) and (\ref{l3}) become:
    \begin{equation}
    2\frac{df}{d\eta}=e^{2h}+e^{2g}-e^{2f}, 
\end{equation}
\begin{equation}
    2\frac{dh}{d\eta}=e^{2g}+e^{2f}-e^{2h},
\end{equation}
\begin{equation}
    2\frac{dg}{d\eta}=e^{2f}+e^{2h}-e^{2g},
\end{equation}
which can be solved exactly and yield \cite{ghezelbash2008supergravity}:
\begin{equation}
    f(\eta)=\frac{1}{2}\ln(c^2\frac{\textbf{cn}(c^2\eta,k^2)\textbf{dn}(c^2\eta,k^2)}{\textbf{sn}(-c^2\eta,k^2)}),
\end{equation}
\begin{equation}
    h(\eta)=\frac{1}{2}\ln(c^2\frac{\textbf{cn}(c^2\eta,k^2)}{\textbf{dn}(c^2\eta,k^2)\textbf{sn}(-c^2\eta,k^2)}),
\end{equation}
\begin{equation}
    g(\eta)=\frac{1}{2}\ln(c^2\frac{\textbf{dn}(c^2\eta,k^2)}{\textbf{cn}(c^2\eta,k^2)\textbf{sn}(-c^2\eta,k^2)}),
\end{equation}
in terms of the standard Jacobi elliptic functions  $\textbf{sn}$, $\textbf{cn}$ and $\textbf{dn}$ \cite{meyer2001jacobi}.
By changing the coordinate $\eta$ to $r=\frac{2c}{(\textbf{cn}(c^2\eta,k^2))^{1/2}}$, the triaxial Bianchi type IX metric can be written as \cite{ghezelbash2008supergravity}:
 \begin{equation}
     ds^2_{tr. BIX}=\frac{dr^2}{{J(r)}^{1/2}}+\frac{r^2}{4}{J(r)}^{1/2}(\frac{\sigma^2_1}{1-\frac{a^2_1}{r^4}}+\frac{\sigma^2_2}{1-\frac{a^2_2}{r^4}}+\frac{\sigma^2_3}{1-\frac{a^2_3}{r^4}}), \label{biatri}
\end{equation}
where
\begin{equation}
      J(r)=(1-\frac{a_1^4}{r^4})(1-\frac{a_2^4}{r^4})(1-\frac{a_3^4}{r^4}). \label{JJJ}
\end{equation} 
In equation (\ref{JJJ}), $a_i$'s are three constant parameters.
}

\section{The explicit expressions for the Maxwell field equations where $N=4$ }

The Maxwell field equations $M^\phi$, $M^\psi$ and $M^\theta$ are given as,
\begin{eqnarray}
M^\phi&=&\frac{-64}{a}r^2\sqrt{\frac{r^8-16c^4(k^4+1)r^4+256c^8k^4}{r^8}}\cos{\psi}\sin{\psi}\nonumber\\
   &\times& \, c^4\alpha H^EE(\frac{\partial H}{\partial \theta})R^M(\frac{\partial R}{\partial t})(4aW+4Ma+8a),\label{Mt}
\end{eqnarray}
\begin{eqnarray}
    M^\psi&=&\frac{-16}{a}r^2\sqrt{\frac{r^8-16c^4(k^4+1)r^4+256c^8k^4}{r^8}}\cos{\psi}\sin{\psi}\nonumber\\
  &\times& \,  c^4\cos{\theta}\alpha H^EE(\frac{\partial H}{\partial \theta})R^M(\frac{\partial R}{\partial t})(aW+Ma+2a)(k^4-1),\label{Mphi}
\end{eqnarray}
and
\begin{eqnarray}
    M^\theta&=& \frac{-1}{\alpha}[(-16\sin^2{\psi}c^4k^4-16\cos^2{\psi}c^4+r^4) (aW+Ma+2a) \nonumber \\
    &\times& \, (\alpha H^EE(\frac{\partial H}{\partial \theta})R^M(\frac{dR}{dt})]. \label{Mthe}
\end{eqnarray}
The equations (\ref{Mt})-(\ref{Mthe}) lead to the same relation for the constants $W$ and $M$, as $W+M=-2$.

The Maxwell field equation $M^t$ gives a partial differential equation for the metric function $H(r,\theta)$, and is given by,
\begin{eqnarray}
    M^t&=&\frac{1}{4a}(E\alpha R^MH^E(-1024\sin{\theta}H(\frac{\partial H}{\partial r})c^8k^4a+1024r\sin{\theta}(\frac{\partial H}{\partial r})^2c^8k^4a-256\cos^2{\psi}\cos{\theta}\nonumber\\
  &\times &  (\frac{\partial H}{\partial \theta})c^4k^4r^3Ha-64r^5\sin{\theta}(\frac{\partial H}{\partial r})^2c^4k^4a-256\cos^2{psi} (\frac{\partial H}{\partial \theta})^2c^4\sin{\theta}r^3a-64E \nonumber\\
 &\times &   (\frac{\partial H}{\partial r})^2\sin{\theta}c^4r^5a-64r^4\sin{\theta}H(\frac{\partial H}{\partial r})c^4a-36a(\frac{\partial H}{\partial r})^2\sin{\theta}c^4r^5L+256\cos^2{\psi}(\frac{\partial H}{\partial \theta})^2\nonumber\\
 &\times&   \sin{\theta}ac^4k^4r^3+256\cos^2{\psi}H(\frac{\partial H}{\partial \theta})\cos{\theta}ac^4r^3-64r^5\sin{\theta}(\frac{\partial H}{\partial \theta})^2 \nonumber\\
  &\times&  c^4a+12r^8\sin{\theta}H(\frac{\partial H}{\partial r})a+16a(\frac{\partial H}{\partial \theta})^2\sin{\theta}r^7L+16\cos{\theta}H(\frac{\partial H}{\partial \theta})r^7a\nonumber\\
   &+& 16E(\frac{\partial H}{\partial \theta})^2\sin{\theta}r^7a+4E(\frac{\partial H}{\partial r})^2\sin{\theta}r^9a+4a(\frac{\partial H}{\partial r})^2\sin{\theta}r^9L-256E\cos^2{\psi}(\frac{\partial H}{\partial \theta})^2\nonumber\\
 &\times&   c^4\sin{\theta}r^3a-256a\cos^2{\psi}(\frac{\partial H}{\partial \theta})^2\sin{\theta}c^4r^3L+4r^9\sin{\theta}H(\frac{\partial^2 H}{\partial r^2})a\nonumber\\
   & +&16\sin{\theta}H(\frac{\partial^2 H}{\partial \theta^2})r^7a-256\cos^2{\psi}(\frac{\partial^2 H}{\partial \theta^2})c^4\sin{\theta}r^3Ha-256H(\frac{\partial^2 H}{\partial \theta^2})c^4k^4r^3a\nonumber\\
 &+&   1024r\sin{\theta}H(\frac{\partial^2 H}{\partial r^2})c^8k^4a-64r^5\sin{\theta H(\frac{\partial^2 H}{\partial r^2})}c^4k^4a+256a\cos^2{\psi}(\frac{\partial H}{\partial \theta})^2\nonumber\\
  &\times&   \sin{\theta}Lc^4k^4r^4+256\cos^2{\psi}(\frac{\partial H}{\partial \theta})^2\sin{\theta}Eac^4k^4r^3-256a(\frac{\partial H}{\partial \theta})^2\sin{\theta c^4k^4r^3L}\nonumber
  \end{eqnarray}
\begin{eqnarray}
&-&64a(\frac{\partial H}{\partial r})^2\sin{\theta}c^4k^4r^5L+1024a(\frac{\partial H}{\partial r})^2\sin{\theta}c^8k^4rL-64r^4\sin{\theta}H (\frac{\partial H}{\partial r})c^4k^4a\nonumber\\
  &-&  64r^5\sin{\theta}H(\frac{\partial^2 H}{\partial r^2})c^4a-256E(\frac{\partial H}{\partial \theta})^2\sin{\theta}c^4k^4r^3a+1024E(\frac{\partial H}{\partial r})^2\sin{\theta}c^8k^4ra\nonumber\\
    &-&64E(\frac{\partial H}{\partial r})^2\sin{\theta}c^4k^4r^5a+256\cos^2{\psi}H(\frac{\partial^2 H}{\partial \theta^2})\sin{\theta}ac^4k^4r^3-256\sin{\theta}(\frac{\partial H}{\partial \theta})^2\nonumber\\
  &\times&   c^4k^4r^3a-256\cos{\theta}r^3H(\frac{\partial H}{\partial \theta})c^4a+16\sin{\theta}(\frac{\partial H}{\partial \theta})^2r^7a+4r^9\sin{\theta}(\frac{\partial H}{\partial \theta})^2a.\label{Mtt}
\end{eqnarray}
We find that solutions to equation (\ref{Mtt}) is given by,
\begin{equation}
H(r,\theta)=(j_+r^2\cos{\theta}+j_-)^{\frac{2}{a^2+2}},
\end{equation}
where $j_+$ and $j_-$ are two constants of integration.
\end{appendix}
\section*{References}
\bibliographystyle{unsrt}
\bibliography{main}
\end{document}